\documentclass{aa}  
\usepackage{graphics,graphicx,lipsum}
\usepackage{wrapfig}
\usepackage{caption}
\usepackage{subcaption}
\usepackage{amsmath}
\usepackage{amssymb}
\usepackage{color, xcolor}
\usepackage[utf8]{inputenc}
\usepackage{natbib}
\usepackage{enumitem}
\usepackage{xparse}
\usepackage{float}
\usepackage{dblfloatfix}
\usepackage{placeins}
\usepackage{txfonts}
\usepackage{natbib}
\usepackage{scrextend}
\usepackage{hyperref}
\usepackage{cleveref}
\usepackage{tabularx}
\usepackage{booktabs}
\usepackage{array}
\usepackage{rotating}

\def\Hii{H\,{\sc ii}}

\def\msun{$M_{\odot}$}

\def\Msun{$M_\odot$}
\def\Mearth{M$_\oplus$}

\def\micron{$\mu$m}

\begin{document}

   \title{ALMA detections of circumstellar disks in the giant H~{\sc ii} region M17}

   \subtitle{Probing the intermediate- to high-mass pre-main-sequence population}

   \author{\mbox{J. Poorta}\inst{1}
         \and
         \mbox{M. Hogerheijde}\inst{1,2}
         \and
         \mbox{A. de Koter}\inst{1,3}
         \and
         \mbox{L. Kaper}\inst{1}
         \and
         \mbox{F. Backs}\inst{3}
         \and
         \mbox{M.C. Ram\'{i}rez Tannus}\inst{4}
         \and
         \mbox{M. K. McClure}\inst{2}
         \and
         \mbox{A. P. S. Hygate}\inst{2}
         \and
         \mbox{C. Rab}\inst{5,6}
         \and
         \mbox{P.D. Klaassen}\inst{7}
         \and
         \mbox{A. Derkink}\inst{1}
    }

   \institute{
        Anton Pannekoek Institute for Astronomy, University of Amsterdam,
              Science Park 904, 1098 XH Amsterdam, The Netherlands\\
              \email{j.poorta@uva.nl}
        \and
        Leiden Observatory, Leiden University, PO Box 9513, 2300 RA Leiden, The Netherlands
        \and
        Institute of Astrophysics, Universiteit Leuven, Celestijnenlaan 200 D, 3001 Leuven, Belgium
        \and
        Max Planck Institute for Astronomy, Königstuhl 17, 
            D-69117 Heidelberg, Germany
        \and 
       University Observatory, Faculty of Physics, Ludwig-Maximilians-Universität München, Scheinerstr. 1, D-81679 Munich, Germany
        \and
        Max-Planck-Institut für extraterrestrische Physik, Giessenbachstrasse 1, D-85748 Garching, Germany
        \and
        UK Astronomy Technology Centre, Royal Observatory Edinburgh, Blackford Hill, Edinburgh EH9 3HJ, UK
    }

 
  \abstract
 {Our current understanding is that intermediate- to high-mass stars form in a way similar to low-mass stars, that is, through disk accretion. The expected shorter formation timescales, higher accretion rates, and increasingly stronger radiation fields compared to their lower mass counterparts may lead to significantly different physical conditions that play a role in disk formation, evolution, and the possibility of (sub)stellar companion formation therein.}
{We search for the mm counterparts of four intermediate- to high-mass ($4-10$\,\msun) young stellar objects (YSOs) in the giant H~{\sc ii} region M17 at a distance of 1.7~kpc. These objects expose their photospheric spectrum such that their location on the pre-main-sequence (PMS) is well established. They have a circumstellar disk that is likely remnant of the formation process.}
{With the {\it Atacama Large Millimeter/submillimeter Array} (ALMA) we have detected, for the first time, these four YSOs in M17, in Band 6 and 7, as well as four other serendipitous objects. Besides the flux measurements, the source size and spectral index provide important constraints on the physical mechanism(s) producing the observed emission. We apply different models to estimate the dust and gas mass contained in the disks.}
{All our detections are spatially unresolved, constraining the source size to $<120$\,au, and have a spectral index in the range $0.5-2.7$. The derived (upper limits on the) disk dust masses are on the order of a few M$_{\oplus}$ and estimations of the upper limits on the gas mass vary between $10^{-5}$ and $10^{-3}$\,\msun. Our modeling suggests that the inner disks of the target YSOs are dust depleted. In two objects (B331 and B268) free-free emission indicates the presence of ionized material around the star. The four serendipitous detections are likely (low-mass) YSOs. We compare the derived disk masses of our M17 targets to those obtained for YSOs in low-mass star-forming regions (SFRs) and Herbig stars, as a function of stellar mass, age, luminosity, and outer disk radius. The M17 sample, though small, is both the most massive and the youngest sample, yet has the lowest mean disk mass.}
{The studied intermediate- to high PMS stars are surrounded by low-mass, compact disks that likely do not offer a significant contribution anymore to either the final stellar mass or the formation of a planetary system. Along with the four serendipitous discoveries our findings showcase the capability of ALMA to probe disks in relatively distant, high-mass SFRs, and offer tentative evidence of the influence of the massive star formation environment on disk formation, lifetime, and evolution.} 

  \maketitle
%

\section{Introduction} \label{sec:introduction}

Recent developments in instrumentation, and more specifically, the advent of the {\it Atacama Large Millimeter/submillimeter Array} (ALMA), have made it possible to perform population studies of disks around low-mass young stellar objects (YSOs) and T Tauri stars. Studying such disk populations in a statistical manner and linking disk properties to age, exoplanet demographics and stellar parameters sheds light on the evolution of disks, including their dissipation timescales and mechanisms, as well as (sub)stellar companion formation within them \citep[e.g.][]{pascucci2016,ansdell2017,andrews2013,tripathi2017, pinilla2020,manara2023}. Recently, this endeavor has been extended towards higher masses, that is, incorporating Herbig AeBe stars (mostly) in the lower mass range around 2\,$M_{\odot}$, with a few higher mass objects up to  10\,\msun\,\citep{stapper2022}, who conclude that the disks of the studied Herbig stars are skewed towards higher mass and larger size than those of T~Tauri stars.

It remains challenging to push this endeavor to higher masses, as the number of young objects becomes progressively smaller with increasing mass, and the birth environment of such objects is complicated by their large distance, crowding, and extinction. As a result, many of the questions regarding the lifetime and dissipation mechanisms of disks, the mechanism of stellar and, possibly, planetary companion formation remain elusive in the higher mass ranges. A fair number of disks, outflows, and disk-like structures have been observed in mm- and cm-wavelengths around massive ($\geq$\,8\,\msun) young or proto-stellar objects \citep[e.g.,][]{johnston2015,johnston2020,beltran2016,ilee2016,carattiogaratti2017, cesaroni2017,ilee2018a, beuther2018a, maud2018,sanna2019,burns2023}. However, the observed structures extend scales of hundreds to thousands au and, unlike in the quoted disk (population) studies, belong to embedded, early stages of formation when the photosphere of the central object is still obscured and its (final) mass is uncertain \citep[see, however,][]{mcleod2024}. 

We have searched for candidate massive YSOs in star-forming regions (SFRs) associated with ultra-compact H~{\sc ii} regions \citep[e.g.][]{bik2004,hanson1996,lumsden2013}. In this context, a region of particular interest is the M17 H~{\sc ii} region (\Cref{fig:m17_overview}) in the east side of the Omega (M17) giant molecular cloud complex, at a distance of $\sim$\,1.7\,kpc \citep{maizapellaniz2022,stoop2024}. Its ionizing source, the young massive cluster NGC~6618, contains more than 100 B stars and at least 16 O stars. At least four of these stars have spectral types earlier than O6 V, the spectral type of $\theta^1$ Ori C, the most luminous star in the Trapezium region of the Orion Nebula Cluster (ONC) \citep{hanson1997, povich2007, povich2009,hoffmeister2008}. With an estimated total stellar mass of $\sim$\,$8000-10,000$\,\msun, NGC~6618 is about five to six times as massive as the ONC \citep{povich2009,kroupa2018}. An important property of the candidate massive YSOs in this region is that they have visible photospheres, allowing to place them on the Hertzsprung-Russell diagram \citep[HRD;][]{Ochsendorf2011}. 

\citet[hereafter RT17]{ramirez-tannus2017} characterized the young stellar population in NGC 6618 constraining its age to $\lessapprox 1$ Myr. 
Recently, the distance estimate to M17 was adjusted from 1.98 kpc \citep{xu2011} to $\sim$\,1.7\,kpc using {\it Gaia} DR3 data \citep{maizapellaniz2022,stoop2024}.
Also the age estimate has been refined by \cite{stoop2024} who used the kinematic age of dynamically ejected massive runaway stars yielding an age of $0.65\pm0.25$\,Myr. Within NGC 6618, RT17 identified a sample of pre-main-sequence (PMS) stars with masses $\sim$\,$4-15$\,\msun\,that optically reveal their photospheres and, at the same time, show optical and near-infrared (NIR) signatures of a circumstellar disk extending close to the stellar surface \citep[][]{backs2023, poorta2023}. Their stellar parameters were first determined by RT17 and later updated by Backs et al. (2024; in prep.), by fitting the normalized photospheric spectra using the non-LTE stellar atmosphere code {\sc Fastwind} \citep{santolaya-rey1997, puls2005} and a genetic algorithm based fitting approach {\sc Kiwi-GA} \citep{brands2022}. In addition to the stellar properties, extensive modeling of the circumstellar disk features in the optical and NIR (\Cref{sec:target_sample}) suggest that these objects are in the final stages of formation and have small scale gaseous and dusty disks. However, constraints on the existence of extended disk material and its mass are lacking so far. 

The sample of PMS stars in M17 fits the common definition of Herbig Be stars in literature, but the objects have not been included in statistical studies and catalogues of Herbig stars \citep[e.g.][]{vioque2018,brittain2023}, likely due to their distance, extinction, and the fact that they reside in the complicated environment of a massive SFR. Additionally, \citet{stapper2022} study a volume-limited sample out to 450 pc, excluding any targets in M17.

In the context of the dearth of disk studies in the higher stellar mass regime, and in massive SFRs in general, this study aims to take the next step by constraining the extended disk properties of four intermediate to massive PMS stars in M17 with masses $\sim$\,$4-10$\,\msun. With ALMA we have detected, for the first time, the mm components of these YSOs, adding to the handful of similar objects for which ALMA observations have been obtained, such as: MWC~297 (B1.5~Ve, $\sim 10$~\msun) a rapidly rotating massive YSO at a distance of 418~pc \citep{sandell2023,stapper2022} and IRS2 (early B) in NGC2024 at a distance of 450~pc \citep{lenorzer2004a, vanterwisga2020}. 
We adopt similar and refined techniques to constrain the disk mass, radius and structure of our targets. For the first time we apply a multi-wavelength approach to this kind of object, including both photospheric optical to NIR spectra, as well as optical to MIR photometry and the newly obtained ALMA mm-wavelength data. 

In the next section (\cref{sec:data}) we present the M17 targets and the ALMA observations. All four targets are detected, and in the covered fields we identify four additional sources, most likely (low-mass) YSOs. In \cref{sec:methods} we analyse the ALMA data and introduce the methods that we use to measure the disk masses. The mm emission is thermal in nature, that is, free-free radiation and/or thermal dust emission. Different estimates of the disk masses are presented in \cref{sec:results}. We discuss our results in \cref{sec:discussion} and compare these with those obtained for T~Tauri and Herbig stars. In the final section (\cref{sec:conclusions}) we summarize our conclusions. 

\section{ALMA observations of M17 targets} \label{sec:data}

\subsection{M17 target sample} \label{sec:target_sample}
\begin{table*}[ht!]
\footnotesize
\centering
\caption{Stellar and extinction properties derived from quantitative spectroscopy  and optical ($\lambda \lessapprox 1.5 \mu$) spectral energy distribution (SED) fitting \citep{2024A&A...690A.113B}.}       
\begin{minipage}{0.9\hsize}
\centering
\renewcommand{\arraystretch}{1.4}
\setlength{\tabcolsep}{3pt}

\begin{tabular}{cccccccccccc}
\hline
    Source & Sp. Type\footnote{Spectral types adopted from \cite{derkink2024a}.} & $T_{\rm eff}$	  &  $\log g$               &  $M_{\rm ZAMS}$ \footnote{~Zero-age main-sequence (ZAMS) mass of best fit PMS track.} & $M_{\text{spec}}$\footnote{Spectroscopic mass from modeling.} & $\log L/L_{\odot}$ & $R_{\star}$ & $\rm A_V$ &   $\log Q\footnote{Number of hydrogen ionizing photons.}$ & $v_{\text{rot}}$ & $\rm Age_{\rm HRD}$\\
       &          &  K                  &  $\rm cm$\,$\rm s^{-2}$ &  $M_{\odot}$        &  $M_{\odot}$         &                             & $R_{\odot}$            &                              &                       &    km$\rm s^{-1}$   & Myr                        \\  
\hline
   
B275 & B7 III & $12750_{-1250}^{+750}$  & $3.44_{-0.34}^{+0.12}$  & $7.2_{-0.5}^{+0.5}$ & $9.0_{-4.3}^{+1.9}$  & $3.32_{-0.09}^{+0.05}$      & $9.5_{-0.4}^{+0.7}$    &  $7.41_{-0.07}^{+0.07}$             & $42.4_{-3.0}^{+0.26}$ & $185_{-75}^{+65}$    & $0.20_{-0.041}^{+0.054}$   \\
B331\footnote{B331 is too embedded to allow for full modeling. Physical parameters were estimated by RT17, using spectroscopic classification and estimating $T_{\rm eff}$ and $L$ from Kurucz calibration tables. The RT17 values were adjusted to the latest distance estimate of $1.7~ \rm kpc$ (see \Cref{sec:target_sample}).} & late-B &   13000                 &  $-$                    & 10                  &  $-$                 & $3.97^{+0.37}_{\downarrow}$ & $18.7^{+9.6 }_{-7.2}$  &  $13.3^{+0.9}_{-0.9}$        & $-$                   &   $-$               &  0.02                      \\ 
B243 & B9 III & $11900_{-300}^{+1400}$  & $3.78_{-0.14}^{+0.22}$  & $4.2_{-0.2}^{+0.4}$ & $4.7_{-1.1}^{+1.9}$  & $2.58_{-0.03}^{+0.11}$      & $4.6_{-0.4}^{+0.2}$    &  $7.92_{-0.06}^{+0.06}$             & $41.0_{-1.7}^{+0.88}$ & $140_{-45}^{+95}$    & $0.98_{-0.21}^{+0.17}$     \\
B268 & B9 III & $11300_{-300}^{+900}$   & $3.78_{-0.12}^{+0.22}$  & $4.5_{-0.2}^{+0.3}$ & $7.0_{-1.3}^{+3.2}$  & $2.66_{-0.04}^{+0.08}$      & $5.6_{-0.4}^{+0.2}$    &  $7.51_{-0.07}^{+0.07}$             & $41.0_{-2.3}^{+0.33}$ & $180_{-35}^{+115}$  & $0.74_{-0.12}^{+0.13}$     \\

\end{tabular}
\centering
\end{minipage}
\label{tab:stellar_properties}
\normalsize

\end{table*}

\begin{figure*}[h]
   \includegraphics[width=1.\hsize]{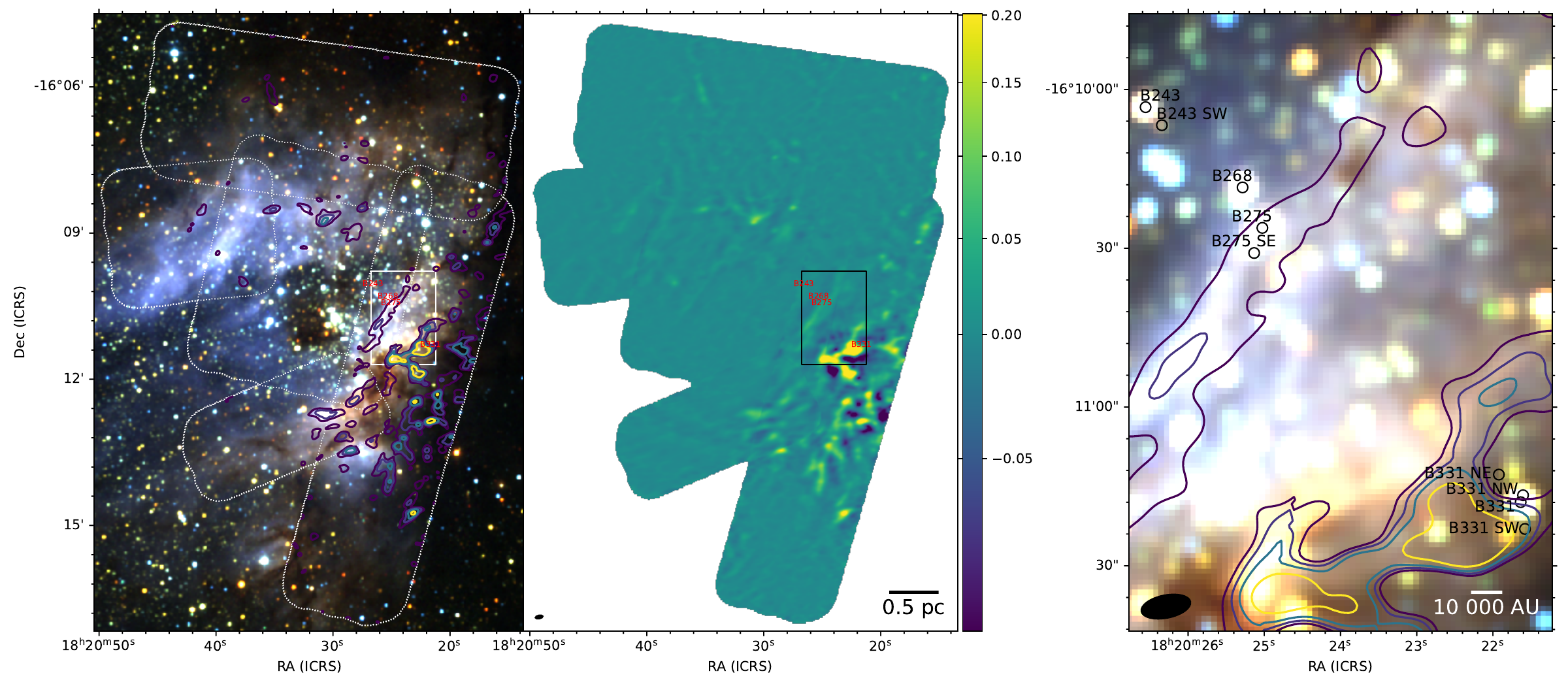}
      \caption{The left panel shows a NIR color composite image of M17 based on 2MASS data: J (blue), H (green), and K (red), with large scale 1.3 mm continuum emission contours (archival ALMA Band 6 data, project code 2018.1.01091.S). Contour levels: 5, 15, 30, 70 $\times$ noise $rms$ (resp. 0.018, 0.053, 0.11, 0.25 Jy). The middle panel shows only the ALMA mosaic, with the beam size in the bottom left corner. On both panels the rectangular region indicates the field in which all four targets in this study (indicated in red) are found. The rightmost panel is a zoom-in on this region with all targets and serendipitous discoveries marked in black. To give an impression of the scales, the radius of each object dot is $1''$, which corresponds to five times the image size of each panel in \Cref{fig:all_detected_sources}.}
         \label{fig:m17_overview}
\end{figure*}

Four of the M17 PMS stars studied by RT17 were observed with ALMA. These four targets are listed in \Cref{tab:stellar_properties} along with their stellar parameters \citep{2024A&A...690A.113B}. Several recent studies have been performed to characterize the inner disks around (some of) these objects. \cite{backs2023} have studied the double-peaked hydrogen emission lines in VLT/X-shooter spectra of B243 and B331 using the radiation thermo-chemical code {\sc ProDiMo} \citep[e.g.,][]{woitke2016}. They find that these lines form very close to the star, such that the disks likely continue (almost) up to the stellar surface, indicating the \citep[expected, e.g.,][]{alecian2013} lack of magnetic fields that are typical for low-mass stars. Based on the line forming region they estimate inner disk ($<20 \, \rm au$) masses on the order of $10^{-3}-10^{-4}$ \msun. They also find that line luminosity-accretion relations derived for low- to intermediate-mass stars \citep{fairlamb2017}, imply accretion rates of $10^{-5}$ to $10^{-3}$~$M_{\odot}$yr$^{-1}$. This would entail a substantial mass reservoir further out in the disk or in-falling onto the inner disk, that, given the evolutionary state of the stars, seems unlikely. 

\cite{derkink2024a} have looked at spectroscopic and photometric variability over days to years in multi-epoch data of B268, B275, and B243. They find intrinsic spectroscopic variability for all these objects in the atomic lines originating mostly close to the stellar surface, as well as evidence for asymmetric disk structures in the inner disk, such as spiral arms. They detect no strong photometric variations, (inverse) P Cygni line profiles, or high-velocity components in forbidden emission lines, making the presence of strong accretion (bursts) or jets unlikely. 

Finally, \cite{poorta2023} have modeled the first and second overtone CO bandhead emission in all four targets to constrain densities and temperatures in the inner disk. In line with previous studies, they find that this emission is typical for high (column) densities ($N_{\rm CO} \sim 10^{21} \, \rm cm^{-2}$) and temperatures ($2000-5000$~K), and originates close to the star ($<1~\rm au$), though not as close as the previously mentioned hydrogen emission lines. They also fit near- to mid-infrared photometry with a dust disk model and find that for a canonical dust-to-gas ratio the derived column densities in the inner disk are inconsistent with those derived from the CO emission (see also \Cref{sec:P23_model_description}). The inner disk structure, though critical for constraining the mass-accretion rate, provides little information about global properties of the disk, such as total mass and size. To investigate these properties, longer wavelength data are needed. 

In \Cref{fig:m17_overview} we present an overview of the M17 region in 2MASS colors and large scale mm emission in ALMA Band 6 from archival data (project code 2018.1.01091.S). The targets in this study are marked on each panel and the right panel shows a close-up of the region containing all four targets, as well as the new detections with ALMA that were made in the field of view around three of the four targets (see \Cref{sec:detections_fluxes}). Some new detections (B243 SW and B275 SE) clearly have 2MASS counterparts. The figure illustrates the high source density as well as the large-scale emission in the region. This emission is largely resolved out in the high-resolution observations taken for this study (\Cref{tab:observational_setup}). \Cref{fig:all_detected_sources} shows a zoom in on all detected sources (see also \Cref{sec:detections_fluxes}).

\subsection{ALMA observations and imaging} \label{sec:alma_observations}
\begin{table*}
\footnotesize
\centering
\caption{ALMA configuration and observational setup per band.}        
\begin{minipage}{0.8\hsize}
\centering
\renewcommand{\arraystretch}{1.4}
\setlength{\tabcolsep}{3pt}
\begin{tabular}{lcc}
\hline
Parameter & Band 6 & Band 7 \\      
\hline
Observing dates (2021) & 23-08 and 30-09  & 7-08 and 14-08 \\
\# 12m antennas & 44 and 45 resp.  &  42 \\
Continuum sensitivity (mJy/beam)    & 0.033 & 0.07 \\ 
Major axis (arcsec)\footnote{See \Cref{tab:gaussian_fitting_results_1} and \Cref{tab:gaussian_fitting_results_2} for details per object.} & 0.06-0.07 & 0.06-0.08\\
Minor axis (arcsec)$^a$ & 0.03-0.04 & 0.04 \\
Position angle (deg)$^a$ & $-$ 73-77 & $-$ 60-64 \\
Maximum recoverable scale (arcsec)     & 0.81 & 0.75 \\
Field of view (arcsec)        & 25.83 & 17.1 \\ 
Integration time (s)      & 1415  & 1100-1200 \\
L5 BL\footnote{Length that includes the 5$^{\rm th}$ percentile of all projected baselines calculated from the unweighted UV distribution.} (m) & $\sim370$ & $\sim270$ \\
L80 BL\footnote{Length that includes the 80$^{\rm th}$ percentile of all projected baselines calculated from the unweighted UV distribution.} (m) & $\sim4500$  & $\sim2900$ \\
Shortest baseline (m) & $\sim47$  &$\sim70$ \\
Longest baseline (m) & $\sim12357$ & $\sim8283$ \\
\end{tabular}
\end{minipage}
\label{tab:observational_setup}
\normalsize
\end{table*}
\begin{table}[htb!]
\footnotesize
\centering
\caption{Spectral setup Band 6.}        
\begin{minipage}{0.9\hsize}
\centering
\renewcommand{\arraystretch}{1.4}
\setlength{\tabcolsep}{3pt}
\begin{tabular}{ccccc}
\hline
Spw & Transition &  Freq\footnote{Transition frequency in the case of lines, central frequency of the spectral window in the case of continuum.}  & \# Chans &  Total BW\footnote{Total bandwidth.} \\
    &            & (GHz) &            & (MHz)     \\
\hline
0 & $^{12}\text{CO}$: v=0 2-1 & 230.538 & 466 & 114 \\
1 & $\text{SO}_2:37(10,28)-38(9,29)$ & 230.965 & 466 & 114 \\
2 & $\rm H(30)\alpha$ & 231.901 & 945 & 231 \\
3 & continuum & 234.375 & 1915 &  1870 \\
4 & continuum & 216.487 & 1915 &  1870 \\
5 & $\text{C}^{18}\text{O}$: v=0 2-1 & 219.560 & 946 &  231 \\
6 & $^{13}\text{CO}$: v=0 2-1 & 220.399 & 946 &  231 \\

\end{tabular}
\end{minipage}
\label{tab:spectral_setup_band6}
\normalsize
\end{table}
\begin{table}[htb!]
\footnotesize
\centering
\caption{Spectral setup Band 7.}        
\begin{minipage}{0.9\hsize}
\centering
\renewcommand{\arraystretch}{1.4}
\setlength{\tabcolsep}{3pt}
\begin{tabular}{cccccc}
\hline
Spw & Transition &  Freq\footnote{Transition frequency in the case of lines, central frequency of the spectral window in the case of continuum.}  & \# Chans &  Total BW\footnote{Total bandwidth.}  \\
    &            & (GHz) &            & (MHz)     \\
\hline
0 & continuum & 346.978 & 238 &  232 \\
1 & $^{12}\text{CO}$: v=0 3-2 & 345.796 & 958 & 234 \\
2 & continuum & 348.358 & 1918 &  1873 \\
3 & continuum & 334.479 & 1918 &  1873 \\
4 & continuum & 336.357 & 1918 &  1873 \\
\end{tabular}
\end{minipage}
\label{tab:spectral_setup_band7}
\normalsize
\end{table}

\begin{figure*}[p]
\vspace{-0.5cm}
   \includegraphics[width=0.9\hsize]{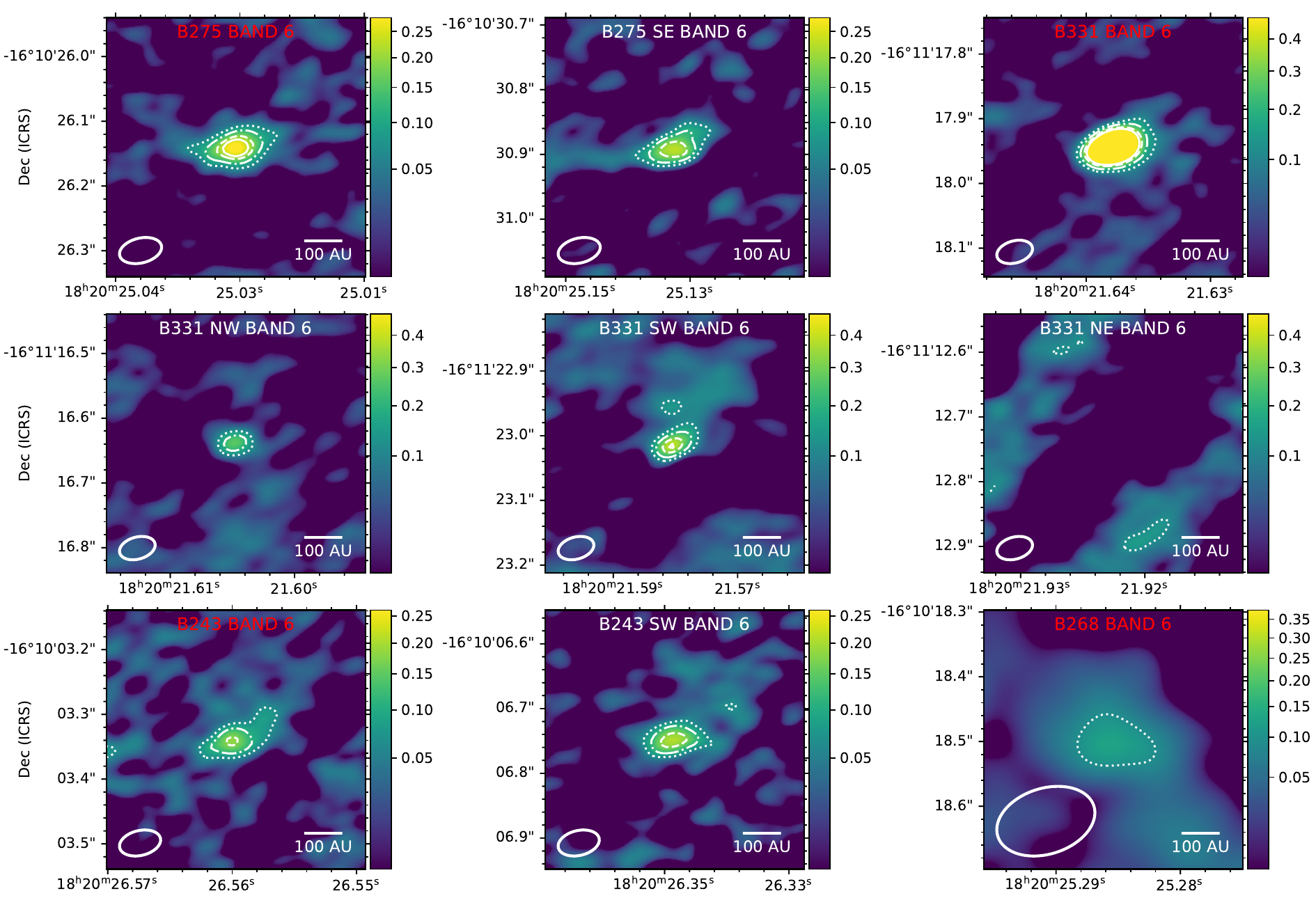}
 \includegraphics[width=0.9\hsize]{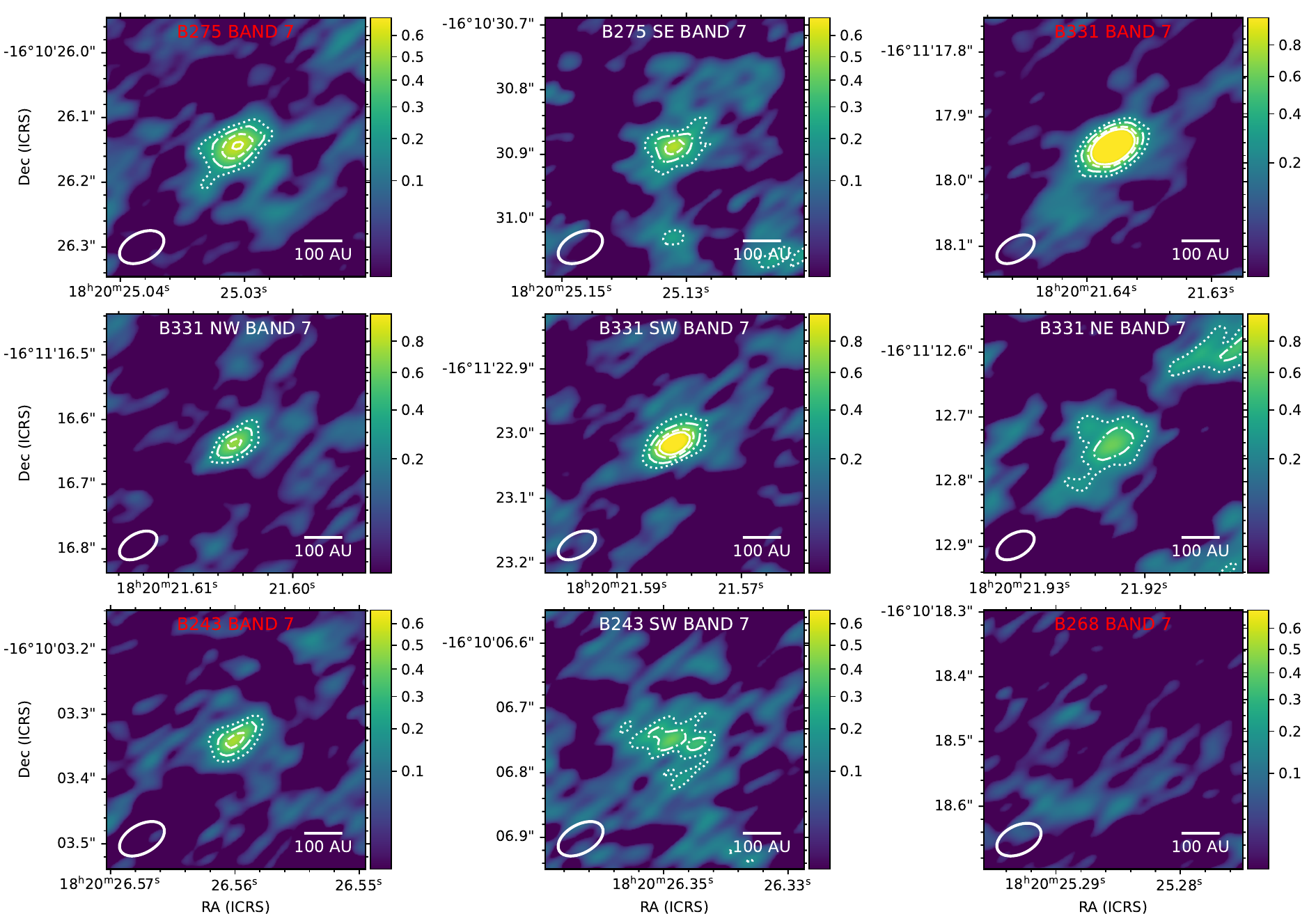} 
\caption{Zoom in on all detected sources. The target sources are labeled in red and the serendipitous detections in white; the latter named after the target in whose image they are detected (see \Cref{sec:detections_fluxes}). Note the slightly different flux scales of each image. Contour levels: 3, 5, 8, 10 $\times$ noise $rms$ (between $0.022-0.040 \rm\,mJy$ in Band 6 and $0.056-0.083 \rm \,mJy$ in Band 7). No radio continuum flux is detected for B268 in Band 7 and B331 NE in Band 6; they still are included as illustration.}     
\label{fig:all_detected_sources}
\end{figure*}

Under project ID 2019.1.00910.S we obtained ALMA 12\,m array observations in Band 6 and Band 7 for our four targets in M17, resulting in eight fields to be imaged. The observational setups for the targets were close to identical; an overview of the observations per band is given in \Cref{tab:observational_setup}. Specific details per target are provided in \Cref{tab:observational_details} in \Cref{sec:obs_details}. The observations were designed to detect the targets in the continuum, but also included the $^{12}\text{CO}$, $^{13}\text{CO}$, and $\text{C}^{18}\text{O}$ rotational 2-1 lines in Band 6 and $^{12}\text{CO}$ rotational 3-2 line in Band 7. The respective spectral setups are summarized in \Cref{tab:spectral_setup_band6,tab:spectral_setup_band7}. The Band 7 data were manually calibrated by the ALMA staff using CASA version 6.1.1 \citep{casateam2022}. The Band 6 data were calibrated with the ALMA pipeline version 2021.2.0.128 using CASA version 6.2.1.7. The signal-to-noise of the detected targets was too low for self-calibration. For all subsequent image deconvolution (imaging) and flux determination we used CASA version 6.2.1.7.

We imaged the continuum using natural weighting to maximize sensitivity, averaging over all spectral windows after including only line-free regions. Because the observed targets lie in a very young SFR, in some cases we had to exclude the data from shorter baselines that are sensitive to extended emission, in order to improve the signal to noise of the observed targets. In Band 6 we used $uv$-ranges of >\,50\,k$\lambda$, thereby excluding emission from scales larger than $\sim$\,$4.13''$, which at the distance of M17 corresponds to $\sim$\,7000\,au. In Band 7 the $uv$-range was already limited to >\,50\,k$\lambda$ when all baselines were included, and we applied no further limit. An exception was made in both bands for the images surrounding B331. As can be seen in \Cref{fig:m17_overview}, this target is located in a region with continuum emission on $\gtrsim$\,$0.4''$\,(700 au) scales, possibly associated with remnant cloud emission or a chance alignment, and we removed baselines <\,500\,k$\lambda$ for images in both bands, thereby excluding emission from scales larger than $\sim$\,700\,au. A zoom-in of the images around each detected source (see \Cref{sec:detections_fluxes}) is shown in \Cref{fig:all_detected_sources}. Larger field images showing the background and multiple sources at once are presented in \Cref{sec:app_field_images}.

In an attempt to detect the targets in the selected CO lines we inspected image cubes averaged to  $1-2~\rm km\,s^{-1}$. We found that all line emission was dominated by large scales, likely from the surrounding cloud. Removing shorter baselines did not yield detections of the continuum-detected sources in $^{12}\rm CO$ nor in $^{13}\rm CO$. In order to estimate upper limits on the gas mass of possible disks (see \Cref{sec:gas_mass_upper_limit}) we determined the root-mean-square ($rms$) of the noise in the line images with $uv$-ranges of $>$\,500\,k$\lambda$ (removing emission from scales larger than $\sim$\,700\,au). Assuming a FWHM line width of $10~\rm km\,s^{-1}$, we measure the noise $rms$ in a cube with $10~\rm km\,s^{-1}$ wide channels to be $\sim$\,0.8\,$\rm mJy\,beam^{-1}$, translating to a $3\sigma$ upper limit to the integrated line flux of $\sim$\,24.0\,$\rm mJy\,beam^{-1}\,km\,s^{-1}$ for the $^{12}\rm CO$ ($\varv=0$;\,$J$\,=\,2$-$1) line on scales $\lesssim 700$\,au.

\subsection{Detections and determination of fluxes in the continuum} 
\label{sec:detections_fluxes}

\begin{table}
\footnotesize
\centering
\caption{Position of the four targets and separation with respect to the newly detected sources.}       
\begin{minipage}{1\hsize}
\centering
\renewcommand{\arraystretch}{1.4}
\setlength{\tabcolsep}{3pt}
\begin{tabular}{ccccccc}
\hline
Obj       & RA              & $\sigma$\footnote{Error on the position.\label{fref:sigma}}  & DEC             & $\sigma$\footref{fref:sigma} & SEP\footnote{The projected distance between the original target and the new detection.}    \\
          & (hh:mm:ss)      & (mas)     & (deg:mm:ss)     & (mas)    &        \\
\hline          
B275      & $18:20:25.0274$ & 3         & $-16.10.26.143$ & 2        &        \\
B275 SE     & $18:20:25.1348$ & 4         & $-16.10.30.889$ & 2        & $5.0''$ \\
\hline                                                                        
B331      & $18:20:21.6372$ & 0.6       & $-16.11.17.946$ & $<1$     &        \\
B331 NW     & $18:20:21.6062$ & 2         & $-16.11.16.638$ & 1        & $1.38''$\\
B331 SW     & $18:20:21.5803$ & 2         & $-16.11.23.014$ & $<1$     & $5.2''$\\
B331 NE     & $18:20:21.9234$ & 5         & $-16.11.12.741$ & 3        & $6.6''$\\
\hline                                                                        
B243      & $18:20:26.5598$ & 6         & $-16.10.03.339$ & 3        &       \\
B243 SW     & $18:20:26.3455$ & 7         & $-16.10.06.749$ & 1        & $4.6''$ \\
\hline 
B268      & $18:20:25.2872$ & 9         & $-16.10.18.497$ & 6        &        \\
\end{tabular}

\centering

\end{minipage}
\label{tab:source_positions_separations}
\normalsize

\end{table}
\begin{table*}[ht!]
\footnotesize
\centering
\caption{Flux and spectral index for all ALMA detections in Band 6 and 7. One object is also detected in VLA observations.}       
\begin{minipage}{1\hsize}
\centering
\hspace{-1.08cm}
\renewcommand{\arraystretch}{1.4}
\setlength{\tabcolsep}{3pt}
\begin{tabular}{c|cccc|cccc|cc}
\cline{2-9} 
            & \multicolumn{4}{c|}{Band 6}                                        & \multicolumn{4}{c|}{Band 7}                                           &                     & \\
\cline{2-9}                                                                                                                                         
Obj         & UV-range   & RMS\footnote{Root-mean-square of the image noise.}     & flux              & M$_{\rm dust}$\footnote{\label{fref:mdust}Dust mass estimate based on \Cref{eq:mdust} assuming that the flux in this band is purely from dust emission.}            & UV-range   & RMS     & flux                & M$_{\rm dust}$ \footref{fref:mdust}               & VLA flux\footnote{Flux in the VLA X-band (10 GHz) according to \cite{yanza2022}. For non-detections an upper limit is given.}            & $\alpha $\footnote{Spectral index between ALMA bands 6 and 7, given by \Cref{eq:alpha}.}       \\
            &($\rm k\lambda$)& (mJy)   & (mJy)             & \Mearth                   &($\rm k\lambda$)& (mJy)   & (mJy)               & \Mearth                   & (mJy)               &                  \\
\hline                                                                                                                                                                    
B275        & $>50$      & 0.0232  &$0.279  \pm 0.028$ & $2.2 \pm  0.2 $    & all        & 0.0575  & $0.573 \pm 0.058$   &  $1.3 \pm 0.1 $    & $<0.025$            & $1.7   \pm 0.5$    \\
B275 SE       &            &         &$0.250  \pm 0.025$ &   $2.3\pm  0.2 $  &            &         & $0.448 \pm 0.058$   &   $1.2 \pm 0.2 $   & $<0.025$            & $1.4   \pm 0.6$    \\
\hline                                                                                                                                                                       
B331        & $> 500$    & 0.0395  &$1.485  \pm 0.149$ & $8.1 \pm  0.8 $    & $> 500$    & 0.0831  & $1.846 \pm 0.185$   &  $2.9\pm 0.3$     & $0.286 \pm0.022$\footnote{\cite{rodriguez2012} report VLA fluxes of $0.46 \pm 0.10 \rm~and~0.66\pm 0.07\rm~ mJy$ at 4.96 GHz and 8.46 GHz respectively.}    & $0.5 \pm 0.5$\footnote{The spectral index between ALMA Band 6 and VLA 10 GHz is $\alpha = 0.53 \pm 0.06$. \cite{rodriguez2012} report an index between VLA points 4.96 GHz and 8.46 GHz of $\alpha = 0.7 \pm 0.5$. }    \\
B331 NW       &            &         &$0.280  \pm 0.040$ &  $2.5 \pm  0.4 $   &            &         & $0.720 \pm 0.083$   &   $1.9 \pm 0.2$ & $<0.025$            & $2.3   \pm 0.6$      \\
B331 SW       &            &         &$0.410  \pm 0.041$ & $3.7 $ $ \pm  0.4 $    &            &         & $1.117 \pm 0.112$   &  $3.0 \pm 0.3$   & $<0.025$            & $2.4   \pm 0.5$      \\
B331 NE \footnote{Only detected in Band 7, not subjected to further analysis.}   &            &         &$<0.119          $ &                         &            &         & $0.441 \pm 0.060$   &                          & $<0.025$            &                     \\
\hline                                                                                                                                                                                               
B243        & $>50$      & 0.0218  &$0.158  \pm 0.022$ &  $2.0\pm  0.3 $      & all        & 0.0556  & $0.476 \pm 0.056$   & $1.8 \pm 0.2 $    & $<0.025$            & $ 2.7  \pm 0.6$     \\
B243 SW       &            &         &$0.201  \pm 0.022$ &  $1.8 \pm  0.2 $    &            &         & $0.257 \pm 0.056$   &  $0.7 \pm 0.2 $   & $<0.025$            & $ 0.6 \pm 0.8$    \\
\hline                                                                                                                                                                       
B268\footnote{This object was imaged with a UV-taper of $0.08''$ to increase signal-to-noise.}        & $>50$      & 0.0313  &$0.138  \pm 0.031$ &$1.6 \pm  0.4 $     & all        & 0.0577  &  $<0.231~(4\sigma)$ & $< 0.81 $     & $<0.025$            & $< 1.2 $            \\
\end{tabular}

\centering
\end{minipage}
\label{tab:fluxes_indeces_masses}
\normalsize

\end{table*}
The noise $rms$ was determined from a large (radius $\sim$\,$5''$) emission-free region in the images resulting from the procedure described in \cref{sec:alma_observations}. Different such regions were probed in each image and the variation of the $rms$ between them was found to be negligible. A source is considered a detection if its flux is higher than five times the noise $rms$ in one or both bands. Some sources minimally fulfilling this condition on the edge of the primary beam were considered noise peaks after visual inspection. Detections that are not among the original targets are named after the target in whose image they are detected; with upper-case letters appended indicating the position of the detection relative to the original target (e.g., SW for southwest etc.). Three of the four original targets are detected in both bands and four new sources are detected in the fields surrounding the original targets (\cref{fig:all_detected_sources} and \cref{sec:app_field_images}). B268 is only possibly detected in Band 6 (SNR 3), but is included in \Cref{fig:all_detected_sources} and further analysis because it is an original target. This object was imaged with a $uv$-taper of $0.08''$ to increase signal-to-noise, leading to the large beam size relative to the other images.  B331 NE is detected in Band 7 (SNR 6) according to the set criteria; however, it is located in the high noise region on the edge of the primary beam, is not detected in Band 6, and has no infrared counterpart (see Figures \ref{fig:field_img_B331_7} and \ref{fig:acq_B331}). For completeness we therefore present it in \Cref{fig:all_detected_sources} and list in Tables \ref{tab:source_positions_separations} and \ref{tab:fluxes_indeces_masses}, but otherwise exclude it from further analysis. All other detections are in both bands and have optical and/or NIR counterparts. Their exact positions and projected separations are listed in \Cref{tab:source_positions_separations}. 

Fluxes were determined using the imfit task in CASA, which fits a two-dimensional Gaussian to the selected image component (full fit results are provided in \Cref{sec:full_fitting_results}). All resulting major-axis FWHM were at most on the order of the beam size, so we concluded that none of the detected sources is resolved. As expected in that case, the resulting peak fluxes (given in Jy/beam) are all equal within errors to the integrated fluxes (given in Jy). Since the sources are contained in one beam we report only the peak fluxes in units Jy. The errors on the flux are taken to be the maximum of the standard flux calibration error of 10\%, the noise $rms$ and statistical errors from the fit. The resulting fluxes and their errors are listed in \Cref{tab:fluxes_indeces_masses}.

In addition to the ALMA fluxes, for all detected sources additional photometry and VLA radio fluxes were collected from catalogues and tables in literature. For the original targets the found photometric points are listed in \cite{poorta2023}. For the new detections the photometric flux points are listed in \Cref{tab:photometry_new_dets}. VLA fluxes were looked up from two extensive surveys in the M17 region with sensitivity down to 0.005 mJy at a beam size of about $0.04 \rm~ arcsec^2$ \citep{rodriguez2012, yanza2022}. The only detected source is B331, in both surveys. We list a $5\sigma$ upper limit for all other ALMA sources. The values are given in \Cref{tab:fluxes_indeces_masses}. Finally, all recovered flux points are plotted in \Cref{fig:sed_fits}.

All the new sources are also detected in acquisition images in the K-band (central wavelength $\lambda_{\rm C} \sim 2.2$ \micron~ and FWHM $\sim 0.41$ \micron) obtained with the LBT telescope on the $\rm 5^{th}~and~6^{th}$ of July 2023 (A. Derkink, 2023, priv. comm.). On these images (\Cref{fig:acq_B275etc,fig:acq_B331}), the sources are clearly distinct from the original targets and have not moved in the almost two years time since the ALMA observations (\Cref{tab:observational_setup}).

\section{Analysis} \label{sec:methods}

In order to draw conclusions on the measured fluxes and to constrain the source characteristics, we perform different analyses. First, to determine whether we can attribute the measured continuum fluxes to dust emission from a circumstellar disk or to free-free emission from ionized material we determine the spectral index in \Cref{sec:spectral_index} and discuss its interpretation in \Cref{sec:spec_index_results}. Second, under simplifying assumptions we calculate an upper limit on the gas mass from the non-detection of CO lines in \Cref{sec:gas_mass_upper_limit}. Thirdly, we estimate dust masses from the measured fluxes directly in \Cref{sec:mass_estimation_simple}. We then compare the measured fluxes with those obtained in a pre-calculated {\sc ProDiMo} model grid in \Cref{sec:mass_estimation_prodimo}. Lastly, in \Cref{sec:mass_estimation_thin_disk}, we estimate disk masses  using an adapted version of the disk model developed by \citet{poorta2023} to fit CO bandhead emission from the same objects. 

\subsection{Spectral index and origin of detected fluxes} \label{sec:spectral_index}
The slope between two or more different radio regimes is a widely used diagnostic toward characterizing the source of emission in the Rayleigh-Jeans limit. This slope is quantified as the spectral index which, following, e.g., \cite{beckwith1990, tychoniec2020}, is defined as:
\begin{equation} \label{eq:alpha}
    \alpha = \frac{\log(F_{\nu_1}/F_{\nu_2})}{\log(\nu_1/\nu_2)}
\end{equation}
with $F_{\nu_{1/2}}$ the flux at frequencies $\nu_{1/2}$, and $\nu_1 > \nu_2$ such that a positive spectral index indicates a descending slope toward longer wavelengths. For all detected sources we calculate the spectral index between ALMA Bands 7 and 6. For B331 we also determine the index between ALMA Band 6 and the VLA X-band at 10 GHz. The adopted frequencies for these bands are $\nu_7 = 341.4183$ GHz, $\nu_6 = 225.4311$ GHz, and $\nu_{\rm X} = 10$ GHz for Band 7, Band 6, and VLA 10 GHz respectively. For B331 we additionally report a spectral index between the two VLA bands at 4.96 GHz and 8.46 GHz from \cite{rodriguez2012}. All the resulting indices are listed in \cref{tab:fluxes_indeces_masses} and discussed in \Cref{sec:spec_index_results}.

\subsection{Upper limit on the gas mass} \label{sec:gas_mass_upper_limit}

Based on the non-detection of the rotational $^{12}\rm CO$ ($\rm J\,=\,2-1$) line, we estimate an upper limit on the gas mass in the possible disks. Given that the beam sizes and $rms$ values are very similar for the different object images we do this for one average limiting flux value $\sim$\,24.0\,$\rm mJy\,beam^{-1}\,km\,s^{-1}$ (\Cref{sec:alma_observations}). With a beam size of $0.07''\times0.04''$ this limit translates to an antenna temperature of $\sim$\,220\,$\rm K\,km\,s^{-1}$. We used the non-LTE molecular radiative transfer code RADEX \footnote{\href{https://var.sron.nl/radex/radex.php}{https://var.sron.nl/radex/radex.php}} \citep{vandertak2007} to estimate line radiation temperatures, using a kinetic temperature $T_{\rm kin}=100~\rm K$, an $\rm H_2$-density of $10^{10}~\rm cm^{-3}$ leading to LTE conditions, and a line width of  $10~\rm km~s^{-1}$ based on Keplerian velocities at $\sim$~30 to 50 au distance from our sources. Under these assumptions RADEX finds our limiting temperature with a CO column density $N_{\rm CO}\approx4\times10^{17}~\rm cm^{-2}$. The emission at this column density is optically thin ($\tau \approx 0.31$). Integrating over the beam size and assuming a CO abundance of $n({\rm CO})/n({\rm H_2}) = 10^{-4}$, this column density results in an upper limit on the gas mass of $\sim$\,$2 \times10^{-5}$ \msun. 

If the disk is much smaller than the $120\times70$\,au beam, the emission is likely optically thick, and, for an assumed kinetic temperature of 100 K, would have a brightness temperature of $\sim$\,100 K. Such a disk would need to take up less than 20\% of the beam area to remain undetected, corresponding to a disk radius of <\,20-30 au, assuming an inclination in the range of 0-60$^{\circ}$. Under the assumption of such a compact, optically thick disk, the CO column density could in principle be arbitrarily high. However, in the studied objects a constraint is provided by the CO bandhead emission, from which \cite{poorta2023} derive a CO column density at the inner rim of the disk of $N_{\rm CO}\sim 1-8 \times 10^{21}~\rm cm^{-2}$. This density is likely to decline further out in the disk with typical power law exponents between -1 \citep[e.g.,][]{woitke2016} and -1.5 \citep[e.g.,][]{poorta2023}. Hence, to derive an upper limit on the gas mass under these conditions we adopt a constant, average CO column density of $N_{\rm CO}\sim 5 \times 10^{20}~\rm cm^{-2}$ and a CO abundance of $n({\rm CO})/n({\rm H_2}) = 10^{-4}$ throughout a disk of 30\,au. In this case we derive an upper limit on the gas mass of $\sim 6 \times10^{-3}$ \msun. 

\subsection{Dust mass estimation: optically thin approach}
\label{sec:mass_estimation_simple}

Many studies use a simple approximation to estimate dust masses for populations of proto-planetary disks, assuming optically thin dust emission from an isothermal disk \citep[e.g.][]{ansdell2016, cazzoletti2019,stapper2022}. Under this approximation the dust mass in the disk is given by \citep{beckwith1990}:
\begin{equation}\label{eq:mdust}
    M_{\rm dust} = \frac{F_{\nu}d^2}{\kappa_{\nu}B_{\nu}(T_{\rm dust})},
\end{equation}
with $F_{\nu}$ the measured flux at frequency $\nu$, $d$ the distance to the source, $\kappa_{\nu}$ the dust opacity at $\nu$ and $B_{\nu}(T_{\rm dust})$ the Planck function at the assumed dust temperature $T_{\rm dust}$. For the dust opacity we adopt the power law:
\begin{equation}\label{eq:kdust}
    \kappa_{\nu} = \kappa_0 \,(\nu/\nu_0)^{\,\beta},
\end{equation}
with $\kappa_0=10 \rm~cm^2/g$, $\nu_0=1000 \rm~GHz$, and $\beta=1$ from \cite{beckwith1990}. The commonly assumed dust temperature for disks around low mass stars is $T_{\rm dust} = 20\rm~K$ \citep{andrews2005}. However, we an\-ti\-ci\-pate higher temperatures due to the higher luminosities of our stars. Following \cite{andrews2013} and \cite{stapper2022} we use:
\begin{equation}\label{eq:tdust}
    T_{\rm dust} = 25 \left(\frac{L_*}{L_{\odot}}\right)^{1/4} ~~{\rm [K],}
\end{equation}
resulting in dust temperatures between $110~\rm K$ and $246~\rm K$ for the objects in \Cref{tab:stellar_properties}. For the serendipitous detections, for which we do not know the stellar luminosities, we adopt $T_{\rm dust} = 150~\rm K$. 

We list the resulting dust mass for all detected sources in \Cref{tab:fluxes_indeces_masses} and emphasize that not for all of these objects dust emission is the most likely source of the measured flux (see \Cref{sec:spec_index_results}). In \Cref{tab:disk_masses} we list the dust masses under this approximation again only for the original targets with the purpose of comparing the results with the other estimates of disk gas and/or dust mass (see \Cref{sec:disk_mass_results}).

\subsection{Disk mass estimation: {\sc ProDiMo} model grid}
\label{sec:mass_estimation_prodimo}
To guide observations and serve as a reference for a first analysis, we computed a grid of models with an extended version of the radiation thermo-chemical code {\sc ProDiMo} \citep{woitke2009,kamp2017} adapted for disks around hot stars \citep{backs2023}. In these models we used the DIANA-standard model setup with a gas to dust mass ratio of 100 \citep{woitke2016}. The grid consists of $10 \times 10 \times 5$ models with, respectively, total disk masses ranging from $5\times10^{-4}$ to $5\times10^{-1}$ \msun, outer radii from $20~\rm au$ to $500~\rm au$, and inclinations from $15^{\circ}$ to $75^{\circ}$. The central object for this grid was chosen to be B331 (see \Cref{tab:stellar_properties}). The inner radius was set to $11.2~\rm au$, the dust sublimation radius for this object. 

In our current analysis we use the resulting SEDs from these models solely as a reference point to compare with our other mass estimations (\Cref{tab:disk_masses}) and with the SEDs we calculate in \Cref{sec:mass_estimation_thin_disk}. We fitted the model SEDs to the available photometry and ALMA points for each of the four targets in the sample. Before fitting we changed the underlying stellar continuum of each model to match the object in question. The best fit {\sc ProDiMo} models are all the same, with parameters $M_{\rm tot} =5 \times 10^{-4}$ \msun\,(lowest in grid), outer radius $233~ \rm au$\footnote{Because the density declines at higher disk radii, the disk's outer regions may easily fall below the detection limit, especially at low disk masses and high resolution observations. Thus, a large model outer radius can still be consistent with a ‘smaller’ observed source.}, and inclination $75^{\circ}$ (highest in grid). Because the grid models are unable to fit the near- to mid-infrared photometry simultaneously with the ALMA points, and also provide little constraint on the disk mass, we use another disk model to acquire an improved fit of the data as described in the next section. 

\subsection{Dust and gas mass estimation: thin disk modeling}
\label{sec:mass_estimation_thin_disk}
\subsubsection{Model description}
\label{sec:P23_model_description}
It is notoriously difficult to model the inner and outer regions of disks in one self-consistent model. As mentioned in \Cref{sec:target_sample}, several previous studies have constrained the inner disk characteristics of the target sample based on infrared photometry and spectroscopy. In particular, \citet[hereafter P23]{poorta2023} constrain the column density and temperature of the gas in the inner parts of the disk ($\lesssim 1-2~\rm au$) by fitting CO bandhead emission, and characterize the dust emission originating further out ($\lesssim 10-30~\rm au$)  by fitting the near- to mid-infrared SED. In this work, we aim to fit the SEDs including both the infrared photometry and the new ALMA-points, while taking into account the previously obtained constraints on the inner disk. For this we adopt the P23 model, which is a 1D Keplerian disk model that uses power laws for the radial density and temperature structure. It predicts the emission components from the CO gas and the dust assuming constant conversion factors for the $\rm H_2$ to CO abundance and the dust-to-gas mass ratio. For this work we model only the dust emission, with a few adaptations with respect to the original model. For the inner disk column densities we adopt the values resulting from fitting the first and second overtone CO bandheads.

Changes in the dust modeling pertain to the dust opacities and the dust-to-gas mass ratio. Where P23 use single grain-size astronomical silicate opacities from \cite{laor1993}, we adopt the mass absorption coefficients calculated by \cite{ossenkopf1994} for dust in proto-stellar cores with the \citet[i.e., MRN]{mathis1977} size distribution. The tabulated dust opacities are for varying gas densities ($n_{\rm H}=10^5$ to $10^8~\rm cm^{-3}$) and for none, thin, or thick ice mantles on the grains. Due to the relatively hot and dense environments expected for the studied objects, we adopt $n_{\rm H}=10^8~\rm cm^{-3}$ and grains without ice mantles. 

\begin{figure}[t!]
   \includegraphics[width=1.0\hsize]{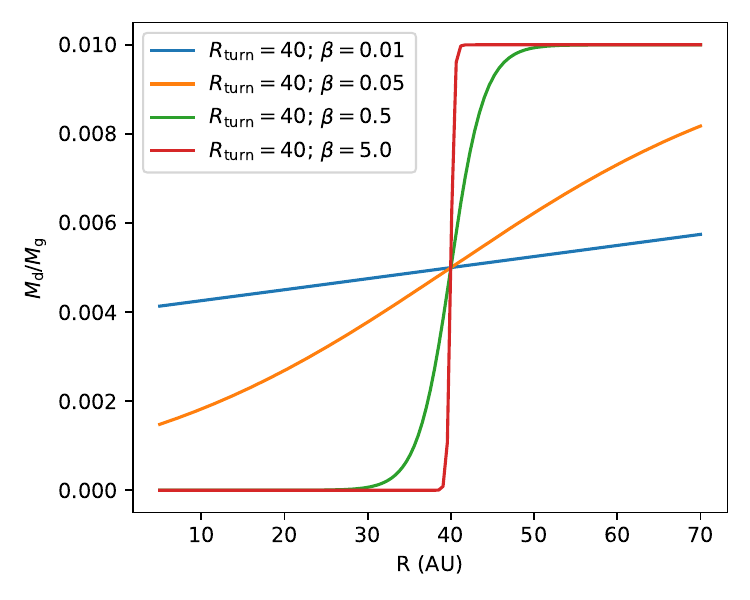}
      \caption{Illustration of the logistic function (Eq. \ref{eq:log_func}) used for varying the dust to gas mass ratio $ M_{\rm d}/M_{\rm g}$ in the disk. The limiting values are fixed to the same values as in the models used for fitting: $(M_{\rm d}/M_{\rm g})_{\rm base}=10^{-8}$ and $(M_{\rm d}/M_{\rm g})_{\rm lim}=10^{-2}$. The value of the free parameter $\beta$ is varied here to illustrate its effect. }
         \label{fig:demo_log_func}
\end{figure}

The second change we introduce is that, instead of using a constant dust-to-gas mass ratio $ M_{\rm d}/M_{\rm g}$ throughout the disk, we vary this quantity using a logistic function as follows (see \Cref{fig:demo_log_func} for an example):
\begin{equation} \label{eq:log_func}
    M_{\rm d}/M_{\rm g} = (M_{\rm d}/M_{\rm g})_{\rm \,base} * \left[ e^{\,\beta*(r-r_{\rm turn})} \right]^{-1} + (M_{\rm d}/M_{\rm g})_{\rm \,lim}.
\end{equation}
The function monotonically rises from the minimum limit $(M_{\rm d}/M_{\rm g})_{\rm base}$ to the maximum limit $(M_{\rm d}/M_{\rm g})_{\rm lim}$ as a function of disk radius $r$ and free parameters $\beta$, the growth rate, and $r_{\rm turn}$, the inflection point. As can be seen in \Cref{fig:demo_log_func}, the function approaches a step function for high values of $\beta$ and a linear function for low values.

The motivation for introducing this function is that the P23 disk models with a constant dust-to-gas ratio cannot account for all the data points at the same time. This also holds for the previously described {\sc ProDiMo} models, which too have a constant dust-to-gas ratio. In both cases, models that fit the longer wavelength points strongly over-predict the infrared fluxes. This is likely connected to the fact that thermal dust emission in the near- to mid-infrared is not a good probe of the dust mass in the inner disk \citep[e.g., P23,][]{vandermarel2023}. This may be due to settling of the bulk of the dust to the mid-plane in the high-density inner regions of the disk, and only a small amount of grains in the disk surface layer contributing to the observable near- to mid-infrared thermal emission. This also leads to a sensitivity of the NIR emission to geometric effects such as the viewing angle. Since the P23 model lacks vertical structure, these effects would not be accounted for. Alternatively, the dust-to-gas ratio could intrinsically vary, due to dust depletion in the inner disk in the presence of, for example, gaps, traps, pebble formation, or other disk structures and processes \citep[e.g.][]{vandermarel2023}. Of course, it is conceivable that a combination of these mechanisms is at play. Regardless of the physical situation, the approach described here is a way to allow the dust distribution in the disk to deviate from the gas column density structure. We briefly discuss interpretations and implications for the total disk mass in \Cref{sec:detection_of_disks}.

\subsubsection{Extrapolation and fitting} \label{sec:mass_estim_extrapol_fitting}
With the P23 model and the previous modeling results we estimate the gas and the dust mass of the disk in two ways. Firstly, without fitting, and not taking into account the ALMA flux points, we directly extrapolate the column density in the inner disk derived from CO bandhead fitting by P23, $(\Sigma_i)_{\rm CO}$, using the power law slope, $q = -1.5$, from the same fits. 
The gas mass is then given by
\begin{equation}
    M_{\rm gas} = \frac{n({\rm H_2})}{n({\rm CO})} \, m_{\rm H} \cdot (\Sigma_{i})_{\rm CO} \, \int_{(R_{i})_{\rm gas}}^{R_{\rm out}} 2\pi r \,  \left( \frac{(R_{i})_{\rm gas}}{r} \right)^{q} \,dr.
\end{equation}
The inner radius for the gas, $(R_i)_{\rm gas}$, is again taken from the CO bandhead fits (P23). The outer radius, $R_{\rm out}$, is set to half the beam size of the image for each object (Tables \ref{tab:fit_results}, \ref{tab:gaussian_fitting_results_1}, and \ref{tab:gaussian_fitting_results_2}). Though B268 was imaged with a $uv$-taper to increase signal to noise (see \Cref{sec:detections_fluxes} and \Cref{fig:all_detected_sources}), for the purposes of this analysis we adopt a similar constraint on $R_{\rm out}$ as for the other sources.
Similarly, the dust mass is given by
\begin{equation}
    M_{\rm dust} = \frac{M_{\rm d}}{M_{\rm g}} \cdot \frac{n({\rm H_2})}{n({\rm CO})} \, m_{\rm H} \, \cdot (\Sigma_{i})_{\rm CO} \, \int_{(R_{i})_{\rm dust}}^{R_{\rm out}} 2\pi r \,  \left( \frac{(R_{i})_{\rm gas}}{r} \right)^{q} \,dr.
\end{equation}
 The inner radius for the dust, $(R_i)_{\rm dust}$, is defined as the point where the dust temperature reaches the dust sublimation temperature $\rm 1500~K$ and is estimated with a radiative equilibrium approach \citep[following][]{lamers1999}, using the dust opacities, the stellar temperature, and stellar radius (\Cref{tab:stellar_properties}). Using the described logistic function (Equation \ref{eq:log_func}) for the dust-to-gas mass ratio has a negligible effect on the total disk mass. Therefore, for the purpose of this extrapolation we assume a constant dust-to-gas mass ratio $M_{\rm d}/M_{\rm g} = 10^{-2}$.

\begin{figure*}[h]
    \begin{center}
   \includegraphics[width=0.9\hsize]{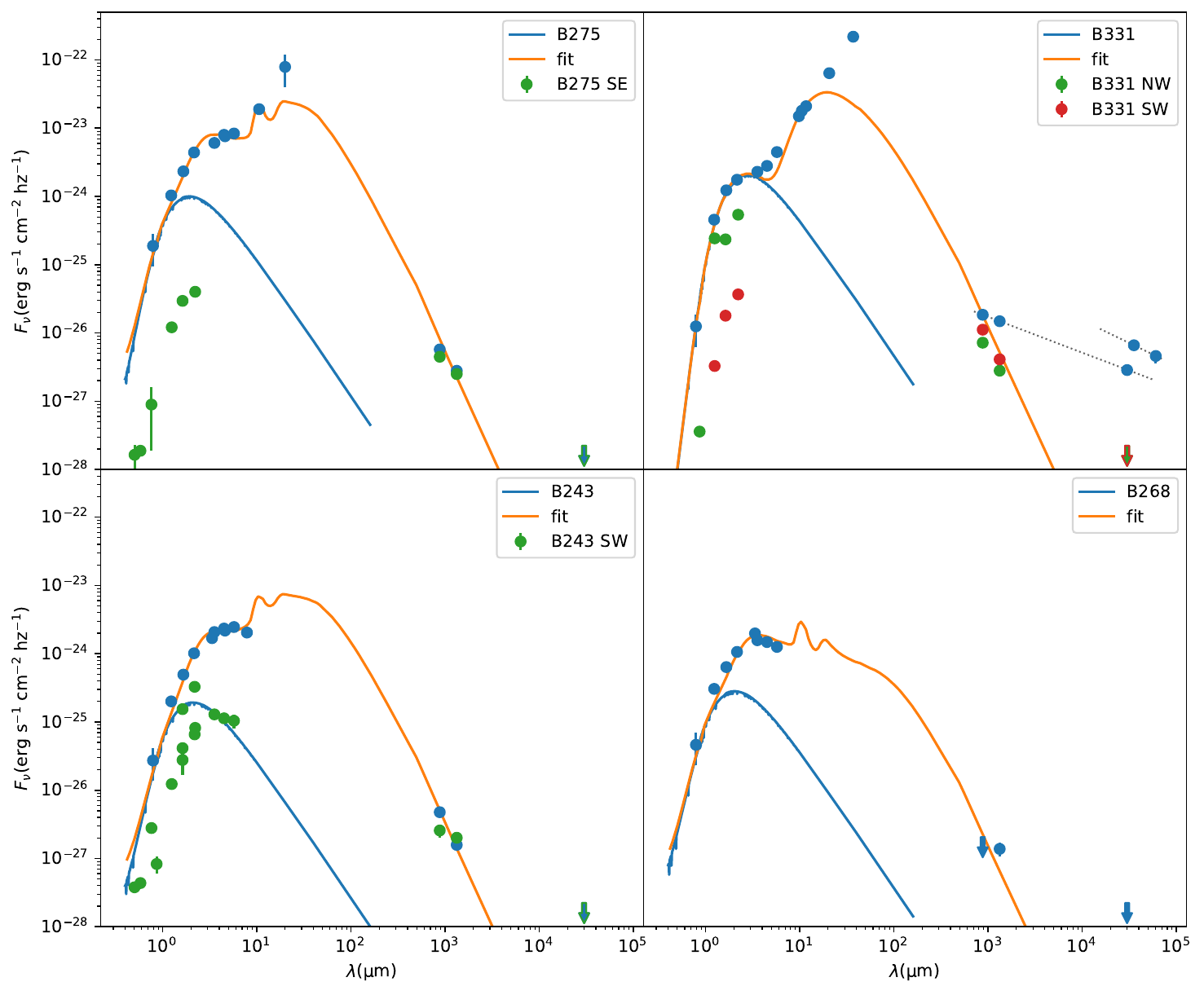}
      \caption{Best fit SEDs, NIR and MIR photometry points, ALMA and VLA flux points for our four original targets and four sources serendipitously detected in the covered fields. Each panel shows the flux data for the original target and the Kurucz model for its stellar continuum in blue. The best fit model SED from the thin disk modeling (described in \Cref{sec:mass_estimation_thin_disk}) is plotted in orange. Each panel also shows the flux data for the serendipitous detection(s) made in the field of the respective original target (in green and red) - these data are not fitted. The downward arrows indicate the upper limit derived for non-detections (mostly VLA 10 GHz). Only B331 (upper right panel) is detected at VLA wavelengths. The dotted lines indicate the slopes between the radio flux points for this object, with a spectral index $\sim 0.67$ for the VLA points at 4.96 GHz and 8.46 GHz \citep{rodriguez2012}, and $\sim 0.52$ between the ALMA and VLA 10 GHz flux points \citep{yanza2022}. There is an offset between the data from \cite{rodriguez2012} and \citep{yanza2022}.}
         \label{fig:sed_fits}
    \end{center}
\end{figure*}

\begin{figure*}[h]
    \begin{center}
   \includegraphics[width=0.9\hsize]{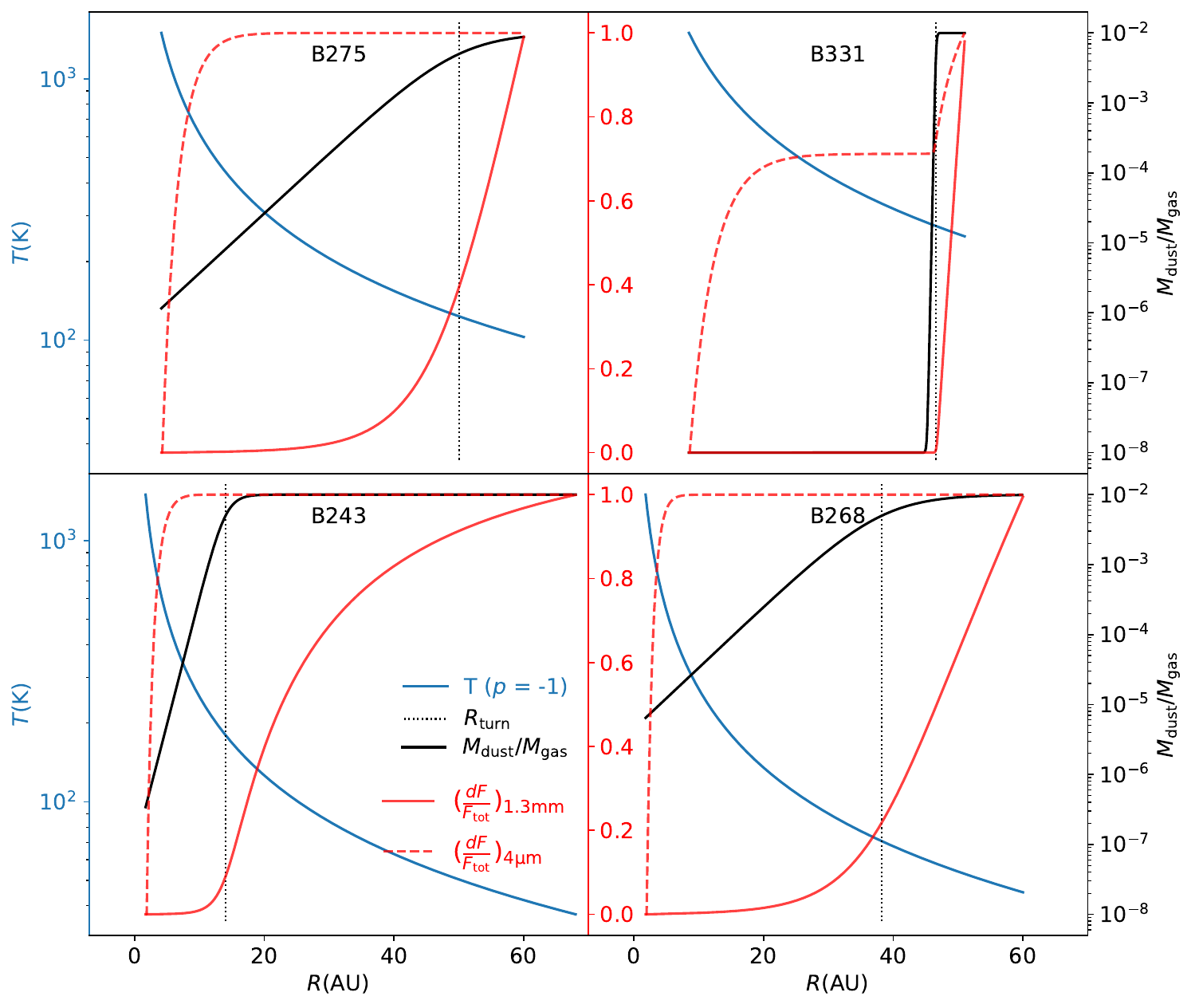}
      \caption{Dust temperature (blue) and dust-to-gas mass ratio (black) as a function of disk radius resulting from the thin disk model fit (described in \Cref{sec:mass_estimation_thin_disk}) for each object. The dotted black vertical line indicates the fitted inflection point $R_{\rm turn}$ of the logistic function of the dust to gas mass ratio. In red is the normalized cumulative flux distribution to illustrate where in the disk the majority of the thermal dust emission originates for different wavelengths and how this relates to the local dust abundance and temperature. The red dashed line is for the flux at $\sim 4$~\micron, the red solid line for the flux at $\sim 1.3$ mm (ALMA Band 6).
      }
         \label{fig:temp_dustgasratio}
    \end{center}
\end{figure*}

Secondly, we fit an SED to the photometry and ALMA flux points using the P23 model with the described adaptations. The SED has a stellar component that is represented by a Kurucz model matching the stellar parameters (\Cref{tab:stellar_properties}). The dust emission from the disk is what we focus on. The fixed parameters going into the modeling are the inner and outer dust disk radius $(R_i)_{\rm dust}$ and $R_{\rm out}$ (defined as before), the inclination $i$ (taken from P23 CO fit results) and a temperature power law exponent which we fix to $p = -1$ based on exploratory fitting. We fit the initial column density $\Sigma_i$ (at $(R_i)_{\rm dust}$), power law exponent $q$, and the logistic function parameters $r_{\rm turn}$ and $\beta$ (Eq. \ref{eq:log_func}). The non-linear least-squares routine {\tt curve\_fit} from the SciPy module \citep[version 1.9.3; ][]{virtanen2020} is used to fit the data. The inverse of the errors on the flux are treated as weights and the errors on the fit parameters are calculated from the covariance matrix returned by the routine. The initial values and bounds of the fit parameters, as well as the fixed parameters and the results are listed in \Cref{tab:fit_results}. The initial value of $\Sigma_i$ and its bounds were determined by extrapolating $(\Sigma_i)_{\rm CO}$ (P23) and its errors to $(R_i)_{\rm dust}$, using $q=-1.5$ (P23). The resulting best fit SEDs are presented together with the data in \Cref{fig:sed_fits}. In \Cref{fig:temp_dustgasratio} we show the temperature and dust-to-gas ratio of the best fit models along with their cumulative flux at two different wavelengths, all as a function of the disk radius. 

The gas and disk masses resulting from the two approaches described in this section (extrapolation and fitting) are listed in \Cref{tab:disk_masses}. These figures and tables are discussed in \Cref{sec:disk_mass_results}.

\section{Results} \label{sec:results}
\begin{table*}[ht!]
\footnotesize
\centering
\caption{SED fit parameters and results (\Cref{sec:mass_estimation_thin_disk}, \Cref{fig:sed_fits})}    
\begin{minipage}{\hsize}
\centering
\renewcommand{\arraystretch}{1.4}
\setlength{\tabcolsep}{3pt}
\begin{tabular}{l|cccc|cccc|cccc}
\hline
\hline
 & \multicolumn{4}{c|}{Fixed parameters}  & \multicolumn{4}{c|}{Initial values and bounds} & \multicolumn{4}{c}{Fit parameters}  \\ 
Name  & $(R_i)_{\rm dust}$  & $R_{\rm out}$  &  $i $\footnote{From P23 CO bandhead fitting results.} & $p$& $\Sigma_i$$^a$ &$R_{\rm turn}$  & $\beta$ & $q$$^a$& $\Sigma_i $ & $R_{\rm turn}$ & $\beta$ & $q$\\
  &  (au) &  (au) &  $(^{\circ})$ & &  $(\rm g~cm^{-2})$ & (au) &  & & $(\rm g~cm^{-2})$ & (au) & & \\
\hline
B275 & $ 4.1 $ & $ 60 $  & $ 30 $ & $-1$& $ 1.6~[0.2, 13]$  & $10~[5, 50]$ & $0.4~[0.01, 10]$ & $-1.5~[-2, -0.1]$ & $ 13 \pm 7 $ & $ 50 \pm 16 $ & $ 0.20 \pm 0.07 $ & $ -1.1 \pm 0.6 $  \\
B331$^{a}$  & $ 8.5 $ & $ 51 $  & $ 30 $ & $-1$& $ 0.5~[0.05,8.3]$ & $10~[5, 50]$ & $0.4~[0.01, 10]$ & $-1.5~[-2, -0.1]$& ($ 8.3^{+13}_{\downarrow}  $ & $ 47 \pm 6$ & $ 10^{\uparrow} $ & $ -0.8 \pm 1.3 $)\\
B243  & $ 1.7 $ & $ 68 $  & $ 50 $ & $-1$& $ 5.8~[0.5, 84]$ & $10~[5, 50]$& $0.4~[0.01, 10]$  &  $-1.5~[-2, -0.1]$& $ 79^{+150}_{\downarrow} $ & $ 14 \pm 1 $ & $ 0.84 \pm 0.22 $ & $ -2 \pm 0.7 $  \\
B268\footnote{Sources dominated by free-free emission.}  & $ 1.8 $ & $ 60 $  & $ 80 $ & $-1$& $ 1.2~[0.4, 3.5]$ & $10~[5, 50]$& $0.4~[0.01, 10]$  &$-1.5~[-2, -0.1]$ & ($ 3.5^{+5.3}_{\downarrow} $ & $ 38^{\uparrow} $ & $ 0.2 \pm 1.2 $& $ -0.8 \pm 3.7 $)  \\
\hline
\end{tabular}
\tablefoot{Arrows up and down indicate lower and upper limits respectively.}
\end{minipage}
\label{tab:fit_results}
\normalsize
\end{table*}
\begin{table*}[ht!]
\footnotesize
\centering
\caption{Disk masses resulting from different methods.}        
\begin{minipage}{1\hsize}
\centering
\renewcommand{\arraystretch}{1.4}
\setlength{\tabcolsep}{3pt}
\begin{tabular}{lccc|cc|cc}
\hline
\hline

& \cref{eq:mdust} Band 6 & \cref{eq:mdust} Band 7 & {\sc ProDiMo} & \multicolumn{2}{c|}{$\Sigma_i = (\Sigma_i)_{\rm CO} $; $q = -1.5$; $ \rm D/G = 10^{-2}$} & \multicolumn{2}{c}{$\Sigma_i$, $R_{\rm turn}$, $\beta$, $q$ from fit} \\ 
Name &  $M_{\rm dust}$ (\Mearth) &  $M_{\rm dust}$ (\Mearth) & $M_{\rm tot}$ (\Msun) & $M_{\rm gas}$ (\Msun) & $M_{\rm dust}$ (\Mearth) & $M_{\rm gas}$ (\Msun) & $M_{\rm dust}$ (\Mearth) \\
\hline

B275 & $2.2 $ $ \pm  0.2 $ & $1.3 $ $ \pm  0.1 $ & $5 \times 10^{-4}$ &  $1.4^{+10}_{-1.2}\times 10^{-4} $  & $3.6^{+26}_{-3.1}\times 10^{-1} $  &  $1.7^{+1.1}_{-0.9}\times 10^{-3} $   & $1.0^{+0.5}_{-0.5} $  \\ 
B331$^{a}$ & ($8.1 $ $ \pm  0.8 $) & ($2.9 $ $ \pm  0.3 $) &  ($5 \times 10^{-4}$) &  $1.1^{+18}_{-1}\times 10^{-4} $  & $2.5^{+38}_{-2.3}\times 10^{-1} $  & ( $2.9^{+4.9}_{\downarrow}\times 10^{-3} $)  & ($1.1^{+1.7}_{\downarrow} $ )  \\ 
B243 & $2.0 $ $ \pm  0.3 $ & $1.8 $ $ \pm  0.2 $ &  $5 \times 10^{-4}$ &  $1.3^{+18}_{-1.2}\times 10^{-4} $  & $4.1^{+56}_{-3.7}\times 10^{-1} $  &  $6.0^{+12}_{\downarrow}\times 10^{-4} $  &  $8.4^{+16}_{\downarrow}\times 10^{-1} $   \\ 
B268\footnote{Sources dominated by free-free emission. Masses derived using ALMA data represent upper limits and are placed between brackets.} &($1.6 $ $ \pm  0.4$) & ($0.81{\downarrow}$) & ( $5 \times 10^{-4}$) &  $3.0^{+5.9}_{-2}\times 10^{-5} $  & $8.7^{+17}_{-5.8}\times 10^{-2} $  &  ($4.6^{+6.9}_{\downarrow}\times 10^{-4} $)  &  ($6.4^{+9.6}_{\downarrow}\times 10^{-1} $) \\

\hline
\end{tabular}
\tablefoot{Disk mass estimates from (from left to right): \Cref{eq:mdust} (columns 1 and 2), {\sc ProDiMo} SED fitting (column 3), extrapolating the CO bandhead fitting results from P23 with disk sizes constrained by ALMA (column 4 and 5), and thin disk model SED fitting (column 5 and 6). The latter two approaches are explained in \Cref{sec:mass_estim_extrapol_fitting}. \\ A discussion and comparison of the listed mass estimates is given in \Cref{sec:disk_mass_results}. For the comparison made to disk dust masses in other disk populations (\Cref{sec:discussion_lmsfr_herbig}) we relied on the dust mass estimates listed in column 2.\\
Arrows up and down indicate lower and upper limits respectively. }
\end{minipage}
\label{tab:disk_masses}
\normalsize
\end{table*}

\subsection{Spectral indices and nature of the sources}
\label{sec:spec_index_results}
The most important diagnostics to distinguish the different physical origins of radio emission are source size and spectral index. We first briefly review the most common emission sources and their typical scales and spectral indices, and narrow down the possible mechanisms for the detected sources. We then consider the spectral indices for the sources separately.

\subsubsection{Origin of radio emission and relation to spectral index}
In general, the physical origin of radio emission can be thermal or non-thermal in nature \citep{yanza2022}. Synchrotron emission by charged particles accelerated in a magnetic field is non-thermal. In a young massive SFR like M17 this could originate from the hot corona's of young low-mass stars \citep[e.g.][]{feigelson1999} or from colliding winds in massive binary systems \citep[e.g][]{dougherty2000}. On much larger spacial scales than considered here, super-shells, for example due to collective stellar winds, may also produce non-thermal emission \citep[e.g.,][]{lozinskaya1999}. Non-thermal emission is most commonly, though not exclusively, associated with a negative spectral index, which we do not observe for any of the sources. For the original targets both mechanisms of compact non-thermal emission are very unlikely. On the one hand, the sources are too massive to harbor a magnetic field associated with a convective envelope. On the other hand, they show no signs of (near) equal mass binarity or strong stellar winds \citep[Backs et al. 2023, in prep.]{ramirez-tannus2017}. Furthermore, the only object massive enough to generate a significant stellar wind would be B331 ($\sim 10$ \msun), but an unresolved companion of similar mass also harboring such a wind would have resulted in a more luminous object. For the new detections a contribution from non-thermal emission cannot be excluded, as these sources could well be low-mass YSOs, but given their spectral indices (\cref{sec:spec_index_results_sources}), it would unlikely be the only contributing mechanism. 

Thermal radio emission can either be due to free-free radiation or thermal dust emission. Since free-free radiation is caused by the (negative) acceleration of charged particles, it is always associated with ionized media. The most common of these that are relevant to the observed region are (remnants of) ultra- or hyper-compact \Hii-regions surrounding young OB stars \citep[e.g][]{avalos2009a, sanchez-monge2011}, ionized winds of massive stars \citep[e.g.][]{snell1986},  externally ionized protoplanetary disks (proplyds) \citep[e.g.][]{stecklum1998, zapata2004}, jets from YSOs \citep[e.g.][]{anglada2018}, and photo-evaporating disk winds \citep[e.g.][]{lugo2004}. Thermal dust emission can originate either from a circumstellar disk or from the surrounding molecular cloud. In the latter case, the emission is more diffuse.

The main question for the original target sample is whether the detected emission can be solely or partially attributed to thermal dust emission from a circumstellar disk, or whether free-free emission should be considered. Since all the detected sources are compact ($\lesssim 150$\,au in diameter), a large-scale origin of the emission, such as diffuse free-free or dust emission from the surrounding \Hii-region and molecular cloud, as well as large-scale jets and outflows can be readily excluded as sources of the measured fluxes. Possible contributors to free-free emission originating close to the star include a small scale \Hii-region\footnote{Estimated sizes of hyper-compact \Hii-regions are between $10^{-3}$ and $10^{-2}$~pc \citep{tielens2005,sanchez-monge2011}, corresponding to $\sim 200-2000$ au.}, the base of a radio-jet, or ionized gas in a stellar wind or a photo-evaporative disk wind. The latter are commonly associated with a spectral index $\alpha \simeq 0.6$ \citep{panagia1975,wright1975,lugo2004}. If all the ionized gas close to the star is optically thin, the emission will have an index $\alpha \simeq -0.1$ in the observed ALMA frequencies. 

Finally, the spectral index for optically thick thermal dust emission is that of a black body, that is $\alpha=2$. For optically thin dust the spectral index will depend on the grain size distribution. Small-sized grains in the interstellar medium (ISM) result in $\alpha^{\rm ISM}_{\rm mm} = 3.7$, whereas lower indices, between 2 and 3, are interpreted as a sign of grain growth in proto-planetary disks around low-mass stars \citep{vandermarel2023}. If the index in such disks is higher, it may indicate a lack of larger grains. 

\subsubsection{Spectral indices for detected sources} \label{sec:spec_index_results_sources}
The spectral index of each detected source is listed in \Cref{tab:fluxes_indeces_masses}. We discuss the targets and serendipitous detections separately. 

For two of the original targets, B243 and B275, we measure a spectral index consistent with a flux dominated by thermal dust emission from a circumstellar disk. The slightly lower index for B275 ($\alpha = 1.7 \pm 0.48$), could signify that there is some contribution from free-free emission, though the upper limit on the VLA 10 GHz measurement suggests that this is minimal (see also \Cref{fig:sed_fits}). For B243 we measure the highest index ($\alpha = 2.7 \pm 0.61$), making it the only target source for which we likely detect (partially) optically thin dust emission, with possibly a particle size distribution depleted in larger grains. The tentative detection of B268 in Band 6 is likely not related to a disk, since the source is undetected in Band 7, giving an upper limit to the spectral index ($\alpha < 1.2$). B331 is the only object also detected in the VLA studies by \cite{rodriguez2012} and \cite{yanza2022}. The flux values in those studies show an offset relative to each other (see \Cref{fig:sed_fits}), but the derived spectral index $\alpha = 0.7\pm 0.5$ by \cite{rodriguez2012} is consistent with our derived index between the two ALMA flux points ($0.52 \pm 0.48$) and also with the index between ALMA Band 6 and VLA 10 GHz ($0.53 \pm 0.06$). These indices clearly indicate that in B331 free-free emission dominates the measured flux, though a thermal dust contribution from a disk cannot be excluded. 

Based on these observations, we assume for our further analysis that the fluxes of B243 and B275 are dominated by thermal dust emission from a circumstellar disk, and those of B268 and B331 by free-free emission. For the latter two objects the (upper limit of) the Band 7 flux was used for fitting. Their fit results and mass estimates based on ALMA data are placed between brackets in \Cref{tab:fit_results,tab:disk_masses}.

Of the serendipitous detections B331 NW and B331 SW have spectral indices well above 2, consistent with (partially) optically thin dust emission from a circumstellar disk. B275 SE ($\alpha = 1.4\pm0.55$) and B243 SW ($\alpha = 0.6\pm0.8$) have indices consistent with a large contribution from free-free emission, though, again, a contribution from dust in a disk cannot be excluded. We discuss the serendipitous detections in \Cref{sec:discussion_serendipitous_discoveries}. 

\subsection{Fit results and disk mass estimates} \label{sec:disk_mass_results}

The best fit values of the thin disk modeling (\Cref{sec:mass_estimation_thin_disk}) are listed in \Cref{tab:fit_results} and the resulting models are shown in \Cref{fig:sed_fits}. The results for B331 and B268 are placed between brackets, because the fitted ALMA Band 7 flux point is either likely dominated by free-free emission (B331) or an upper limit (B268). Fitting simultaneously the short- and long-wavelength data points required a relatively shallow column density decrease ($q>-1.5$, except for B243), and high inner-disk surface densities (all best fit $\Sigma_i$ are on or close to their maximum bound). This reflects that models with a nominal density structure ($q=-1.5$) and a small disk size ($R_{\rm out}\lessapprox 50-70 \rm~ au$) that fit the short-wavelength flux, underpredict the ALMA flux. We found the reverse of this effect in the {\sc ProDiMo} models, which overpredict the short-wavelength fluxes while fitting the ALMA points. 

The course of the dust-to-gas mass ratio in the disk as parameterized by $R_{\rm turn}$ and $\beta$ (\Cref{eq:log_func,fig:demo_log_func}) differs per object. This is visualized in \Cref{fig:temp_dustgasratio}, which shows the dust temperature and best fit dust-to-gas mass ratio in the disk as a function of disk radius. Only for B331 $\beta$ is high, and the dust to gas ratio approaches a step function with $R_{\rm turn}$ indicating where in the disk the dust abundance rises significantly. In all other cases $\beta$ is small, so that the dust abundance rises more or less linearly and $R_{\rm turn}$ indicates where it bends toward its limiting value at $10^{-2}$. \Cref{fig:temp_dustgasratio} also shows the normalized cumulative flux in the NIR (4 \micron) and in ALMA Band 6 (1.3 mm), illustrating where in the disk the emission originates at these wavelengths. From this we see that for B275, B243, and B268 the NIR flux is dominated by hot ($\sim$1000\,K) dust in the inner parts of the disk ($\lesssim 5-10\rm~au$). For B331 there are two contributions: one from very tenuous hot dust in the inner disk, and a second from a small dense ring of cooler dust ($\sim 300~ \rm K$). The ALMA flux mostly originates from dust beyond $R_{\rm turn}$, where the dust abundance is highest and the disk has more surface area. The emitting dust temperatures vary from object to object. For B275 and B331 the best fit models indicate a lack of extended cool ($\lesssim 50-100 \rm~K$) dust, that is, the ALMA flux is dominated by higher temperatures ($\sim 100-200~\rm K$). Since the flux of B268 is an upper limit, these results suggest that only B243 can be said to have an extended disk containing cool dust. 

 \Cref{tab:disk_masses} summarizes all disk mass estimates for the four intermediate- to high-mass PMS stars. Again, the estimates for B331 and B268 that are based on ALMA fluxes are placed between brackets, as the ALMA fluxes for these objects are likely dominated by free-free emission. The estimate based solely on the extrapolation of the inner disk ($\lesssim 1~\rm au$ ) CO bandhead fitting result leads to the lowest gas and dust masses  ($M_{\rm gas} \sim 10^{-5}-10^{-4}~M_{\odot}$). The best fit {\sc ProDiMo} model leads to a mass estimate close to the values from the extrapolation ($M_{\rm tot} = 5 \times 10^{-4}~M_{\odot}$). As this is the lowest-mass model calculated in the grid and it overestimates the NIR emission, this can be considered an upper limit in the context of estimates based on {\sc ProDiMo} modeling. 
 
 The dust masses determined directly from the ALMA fluxes through \Cref{eq:mdust} and from modeling the SEDs with the thin disk model lead to similar dust masses, that are at most an order of magnitude higher than the previous two estimates. The slightly lower dust mass estimate in Band 7 simply reflects how the measured spectral index differs from that assumed for the dust opacity (\Cref{sec:mass_estimation_simple}). The fit results as illustrated in \Cref{fig:temp_dustgasratio} suggest that the inner disks are dust depleted, or, in the case of B331, that there is a dust gap.\footnote{An alternative interpretation of the logistic function that was introduced in \Cref{sec:P23_model_description} will be discussed in \Cref{sec:detection_of_disks}.} In general, gas and dust gaps do not have to coincide \citep[e.g.][]{vandermarel2023}. By varying the dust-to-gas ratio separately from the gas column density, the models allow for the gas to decouple from the dust. If, instead, a gas-dust coupling was enforced, or if the gas would be even more depleted than the dust, this would result in lower gas masses than those presented in column five of \Cref{tab:disk_masses}.  Thus, the gas masses derived from using the adapted P23 model to fit the SEDs can be seen as upper limits. Of course, this does not change the fact that the estimation heavily relies on the adopted results from the CO bandhead fitting. Finally, we recall that the non-detection of CO lines with ALMA suggests an upper limit on the gas mass of $\sim 2 \times 10^{-5}$~\msun~in the optically thin case, or, when assuming an optically thick compact disk, $\sim 6 \times 10^{-3}$~\msun~(\Cref{sec:gas_mass_upper_limit}).

\section{Discussion} \label{sec:discussion}

We present the first ALMA detections of disks around well studied, intermediate- to high-mass PMS stars at distances well beyond 1 kpc. Analysis of their properties reveals that they are mainly low-mass, compact disks, possibly dust-depleted in the inner regions, that likely do not contribute significantly anymore to the final stellar mass, nor to the formation of any significant planetary system (\Cref{sec:detection_of_disks}). Two of our target sources are dominated by free-free emission, indicating the presence of ionized material close to the star (\Cref{sec:discussion_free-free_emission}). We place these findings in the context of trends observed in low mass SFRs and among Herbig stars, and find tentative evidence for an influence of the massive star formation environment on disk formation and evolution (\Cref{sec:discussion_comparison_to_other_populations}). We also report four serendipitous detections, showcasing the capabilities of ALMA to probe disks in high mass SFRs (\Cref{sec:discussion_serendipitous_discoveries}). 

\subsection{Detection of compact mm-emission towards intermediate- to high-mass PMS stars: what have we found?}\label{sec:discussion_whathavewefound}

\subsubsection{Detection of disks} \label{sec:detection_of_disks}
Of the four intermediate- to high-mass PMS stars in M17 presented in \Cref{sec:target_sample}, three were unambiguously detected with ALMA in Bands 6 and 7. B268 was marginally detected in Band 6, providing an upper limit on the Band 7 flux and spectral index. The spectral indices presented in \Cref{sec:spec_index_results_sources} suggest that of the four detections, two (B275 and B243) are likely related to cool dust emission from a disk, while the other two (B331 and B268) have a significant contribution from free-free emission, limiting constraints on possible outer disk masses to upper limits. Different approaches were used to estimate (upper limits to) the dust and gas masses in the disks. For the gas mass, these estimates were partially based on earlier results from fitting CO overtone bandhead emission from the inner disks of the targets. 

The results detailed in \Cref{sec:disk_mass_results} and \Cref{fig:temp_dustgasratio}, obtained by fitting the SEDs with the adapted P23 model, suggest that the inner disks are dust depleted. In \Cref{sec:P23_model_description} it was mentioned that the variation of the dust-to-gas ratio could also reflect a scenario in which the hot surface layers of the inner disk and the geometrical orientation determine the level of NIR flux, rather than the amount of dust itself. The P23 1D (geometrically thin) disk model cannot distinguish between these scenarios (or a combination of them) and if the inner disks are not as dust depleted as the best-fit models suggest, this could increase the dust mass. However, this increase will be minor since the issues only affect the inner part of the disk, while the bulk of the dust still resides in the outer parts. Moreover, the gas masses reported in \Cref{tab:disk_masses} (and the remarks made on the results in \Cref{sec:disk_mass_results}) would remain unaffected. 

The overall picture emerging from the reported results is that the studied objects are surrounded, if at all, by small ($R_{\rm out}\lesssim 50-70 \rm~ au$) circumstellar disks with low dust masses (at most a few \Mearth). Estimations of the gas mass in the disks range from $\sim$\,10$^{-5}$ up to about $\sim$\,10$^{-3}$ \Msun. It is possible, even likely, that the disks are more compact than assumed in the modeling and optically thick in the CO lines as discussed in \Cref{sec:gas_mass_upper_limit}. If that is the case, the gas mass might be as high as $\sim$\,6\,$\times 10^{-3}~$\msun. In any case, the low disk masses imply that all stars in the sample have essentially finished their main accretion phase, that is, have stopped accreting amounts of mass that still could contribute significantly to the final mass of the central star. This is in line with the well-detected photospheres of these objects and the previously reported lack of jets, outflows, or other signs of strong accretion bursts \citep{derkink2024a}. 

In this context the presence of CO overtone bandhead emission in the spectra, which is often associated with accretion and earlier stages of formation \citep[see][and references therein]{poorta2023}, may be surprising. Also, the hydrogen emission lines in the spectra of these targets are often linked to accretion and used to estimate accretion rates in Herbig stars \citep[e.g.,][]{wichittanakom2020}. Accretion rates reported for objects with CO bandhead emission are very diverse, ranging from $10^{-3}$ to $10^{-8}~$\msun$\rm yr^{-1}$ \citep{poorta2023}. Accretion rates of the order $\sim$\,$10^{-7} - 10^{-8}~$\msun$\rm yr^{-1}$ are consistent with the mass (upper limit) estimates.

Our modeling suggests that the inner disks are dust depleted, reminiscent of gap- or ring-like structures in transition disks (TDs) among lower mass PMS stars \citep{vandermarel2023}. Herbig stars are traditionally classified into Group I and Group II disks, originally based on their SED appearance \citep{meeus2001}. The basic characteristics of these groups are summarized in \Cref{tab:Herbig_groupI/II} \footnote{For extensive discussion of the Meeus groups and their properties under different observations see e.g., \cite{garufi2017,brittain2023,vandermarel2023}}. Group I disks can be understood to be the high mass equivalent of T-Tauri TDs, with a large inner cavity, and a mm-bright, extended, massive outer disk component. Classification in terms of Group I/II for our objects is ambiguous. Based on their near- to mid-infrared colors, and their inner dust gap B275 and B331 qualify as Group I. However, in terms of their low mm-brightness and their limited size all objects in our sample qualify as Group II.

\begin{table}[t!]
    \centering
        \caption{Herbig AeBe Group I and II source characteristics.}
    \begin{tabular}{c|c|c|c|c}
         Group &  IR color & NIR size & NIR flux & mm appearance  \\
            & $\left(\frac{F_{30\mu \rm m}}{F_{13\mu \rm m}}\right)$  & (radial) & & \\ [6pt]         
         \hline \\ [-11pt]
          I & high  & extended & bright & inner dust cavity, \\
            &       &           &       & massive outer disk \\   
        \hline \\ [-11pt]
        II & low & compact & faint & faint and compact \\
    \end{tabular}
    \label{tab:Herbig_groupI/II}
\end{table}

We discuss the lack of extended, massive disks in mm-emission further when comparing our results to previously studied disk populations in \Cref{sec:discussion_comparison_to_other_populations}.

\subsubsection{Free-free emission}\label{sec:discussion_free-free_emission}
While some contribution of cold dust to the mm-emission of B331 (and possibly B268) cannot be excluded, the spectral indices indicate a dominant role for free-free emission. The spectral indices found for B331 ($0.5-0.7$) are consistent with a stellar wind or disk wind. However, (a contribution of) other mechanisms such as optically thin free-free emission from ionized gas close to the star or the base of a radio jet cannot be excluded (see \Cref{sec:spec_index_results}). With a mass of $\sim$\,10\,\msun, B331 is the most massive source in the sample. There are no indications of a strong stellar wind in the X-shooter spectra \citep{ramirez-tannus2017}. As can be seen in the SED of B331 (top right panel of \Cref{fig:sed_fits}) it lacks the NIR excess seen in the other sources, the infrared excess only starts at $\sim$\,4-5\,\micron. This translates into the gap-like feature modeled in the dust abundance (\Cref{fig:temp_dustgasratio}), that is, the dust is absent up to $\sim$\,45\,au, after which the dust-to-gas ratio steeply rises to $10^{-2}$. \cite{backs2023} also model this gap and, in the absence of long wavelength data, classify the object as a Group I source. All in all, the spectral index and the derived disk morphology combined could suggest a photo-evaporative disk wind. This is consistent with photo-evaporation mechanisms where the disk is being sublimated further away from the star and a mid-disk gap works its way both out- and inwards \citep{alexander2006a,hollenbach1994}. Similar inner gaps have been observed for massive YSOs by \cite{frost2021a} who also attribute this, at least in part, to photo-evaporation. 

The fact that B268 is only marginally detected indicates that the dust disk of this object is more dissipated than that of the other objects. It is noteworthy that both its stellar characteristics and its infrared SED are very similar to that of B243, for which a mm-disk is most clearly detected (\Cref{tab:stellar_properties,fig:sed_fits}). It is also interesting that the CO bandhead emission for this source is very pronounced \citep{poorta2023}, but that the estimated mass of its disk based on this emission is the lowest of all objects in the sample (columns five and six in \Cref{tab:disk_masses}). This underlines that the (near-)infrared view of a disk does not necessarily predict its properties probed at longer wavelengths. The most likely explanation of the observations of this source remains that its disk is more compact than that of the other, detected sources.

\begin{figure*}[h]
   \includegraphics[width=0.9\hsize]{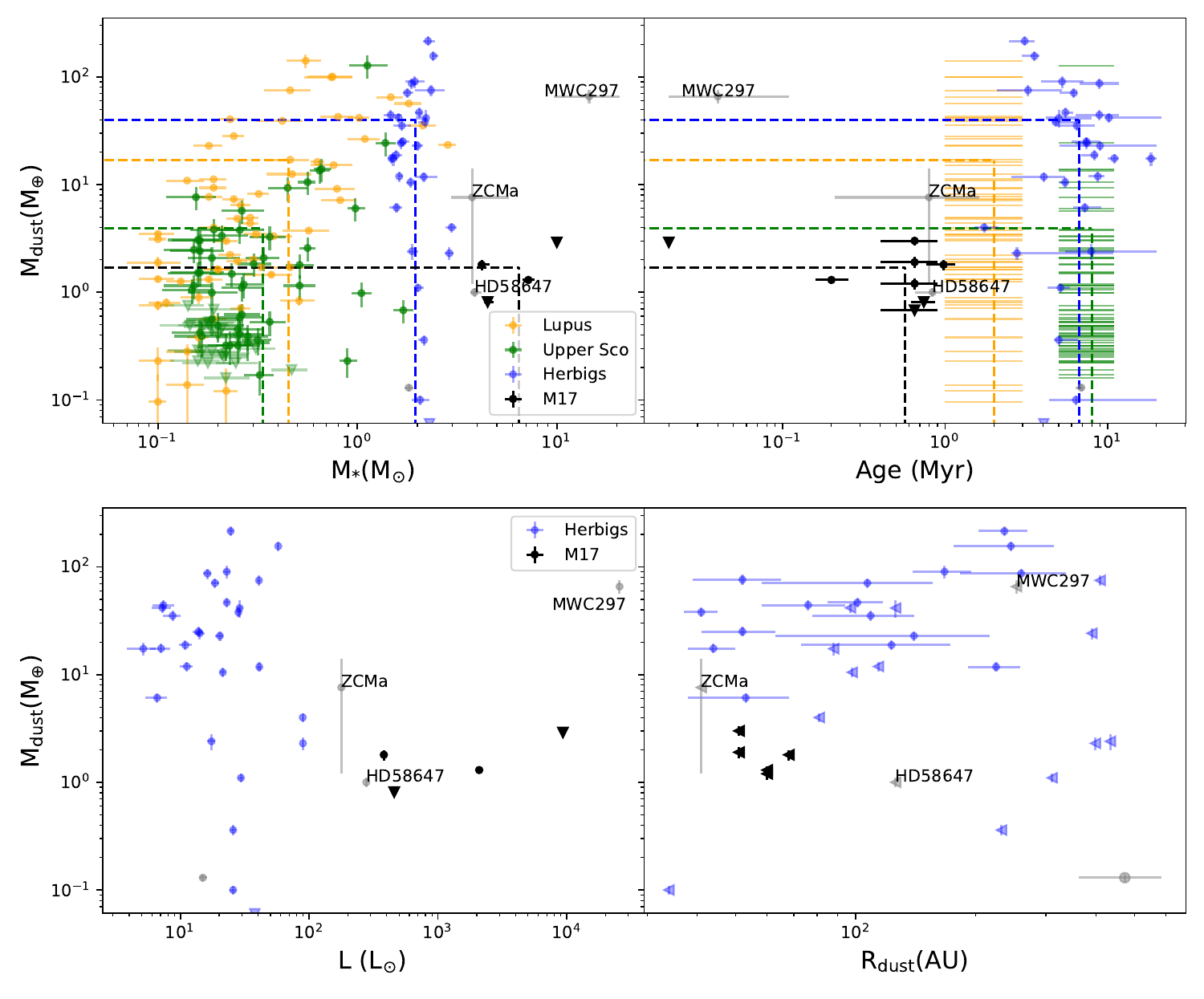}
      \caption{Disk dust mass as a function of stellar mass (upper left), age (upper right), stellar luminosity (lower left), and outer dust radius (lower right), based on ALMA continuum data for different samples. The low-mass PMS populations of the Lupus \citep{ansdell2016} and Upper Sco \citep{barenfeld2016} SFRs are shown in orange and green, respectively. For these populations the age range of the SFR is indicated by plotting each object as a horizontal line in the age-plot. The Herbig Ae/Be sample from \cite{stapper2022} is shown in blue and grey. The grey points are special cases in their sample that are both higher mass and younger - the labeled objects are discussed in the text. In black we show the PMS stars in M17 studied in this work. In the upper panels the mean dust mass, and the mean stellar mass (left) and age (right) of each sample is indicated with horizontal and vertical dashed lines. Upper limits on the dust mass are indicated by downward triangles. Upper limits on the outer dust radius (lower right) are indicated by triangles pointing left.}
         \label{fig:disk_mass_comparison}
\end{figure*}

\begin{figure}[h]
   \includegraphics[width=1.0\hsize]{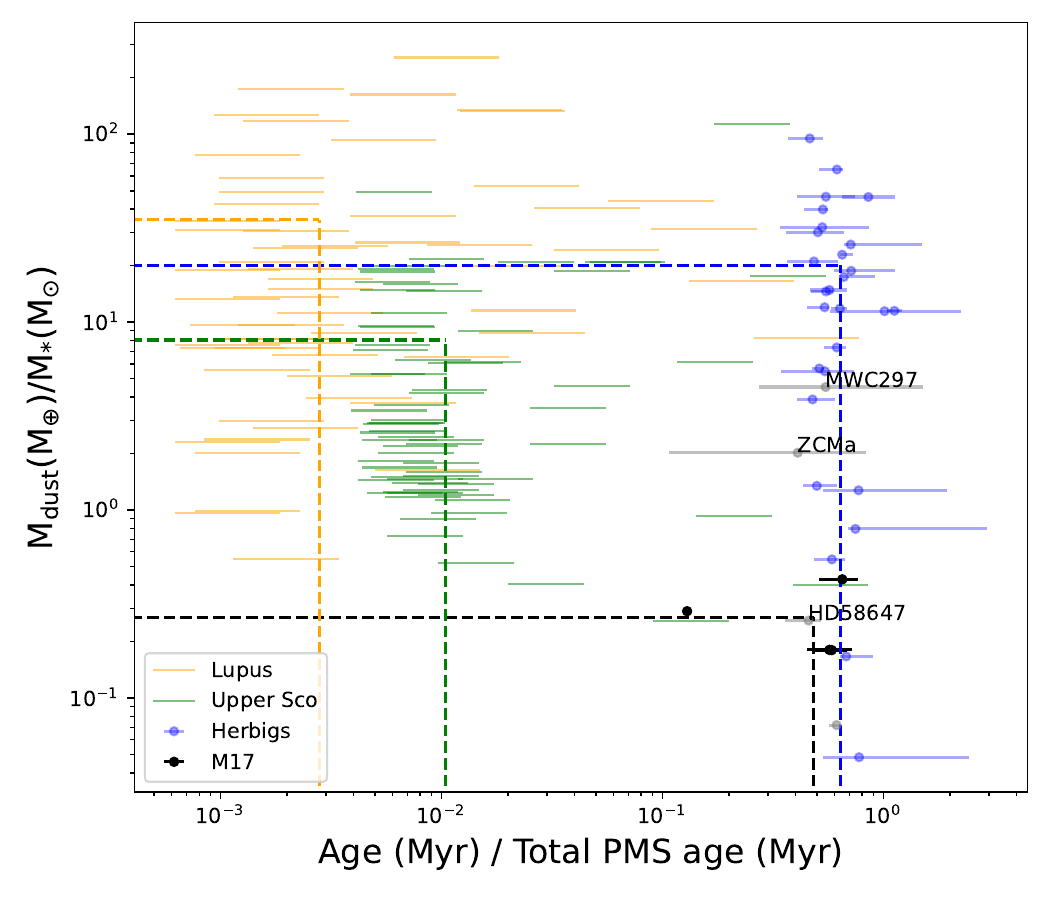}
      \caption{"Normalized" disk dust mass versus "normalized" age for the same samples as in \Cref{fig:disk_mass_comparison}. The dust mass is normalized to the stellar mass of each object. The age is divided by the PMS lifetime of a star of the respective mass, set by the ZAMS-age from the MIST PMS tracks \citep{dotter2016}. For the objects in Lupus (orange) and Upper Sco (green), for which no individual age is known, the age range is indicated by plotting each object as a horizontal line. The dashed lines indicate the mean values of both quantities for the different populations.} 
         \label{fig:normalized_age_vs_dm}
\end{figure}
 
\subsection{Disk masses and sizes compared to other disk populations} \label{sec:discussion_comparison_to_other_populations}

\subsubsection{Low-mass star forming regions and Herbig stars} \label{sec:discussion_lmsfr_herbig}

In \Cref{fig:disk_mass_comparison,fig:normalized_age_vs_dm} we compare our objects with samples from the low-mass SFRs Lupus \citep{ansdell2016} and Upper Sco \citep{barenfeld2016}, and with a sample of Herbig AeBe stars within a distance of $\sim$\,450\,pc studied by \cite{stapper2022}. Dust masses for all the objects in the quoted studies were determined from ALMA Band 6 and/or 7 continuum flux using \Cref{eq:mdust}, be it sometimes with different assumptions on the dust temperature than in this study. For the purpose of comparison we therefore use the dust mass derived from \Cref{eq:mdust}, and because of the (possible) free-free contamination discussed earlier we use the (upper limits on the) dust masses based on the Band 7 flux.

\begin{table}
    \caption{Parameters for M17 objects used in Figures \ref{fig:disk_mass_comparison} and \ref{fig:normalized_age_vs_dm}.}
   \begin{minipage}{\hsize}
\centering
\renewcommand{\arraystretch}{1.4}
\setlength{\tabcolsep}{3pt}
    \begin{tabular}{l|ccccc}
    \hline
\hline
        Object & $\rm M_{dust}$\footnote{Ninth column \Cref{tab:fluxes_indeces_masses}. For objects dominated by free-free emission the values are taken as an upper limit.} & $\rm R_{\rm dust}$\footnote{Upper limit determined by half of beam major axis of the Band 7 images (\Cref{tab:gaussian_fitting_results_2}).} &  $\log L/L_{\odot}$\footnote{From \Cref{tab:stellar_properties}.} & $\rm M_{*}$$^c$  & age$^{c,}$\footnote{For the serendipitous detections the M17 age from \cite{stoop2024} is used.} \\
        & $ \rm (M_{\oplus})$ & au & & ($M_{\odot}$) & (Myr) \\    
        \midrule
        B275 & 1.3$\pm$0.1 & <60 & $3.32^{0.05}_{-0.09}$ & $7.2^{0.5} _{-0.5}$ & $0.20^{0.054} _{-0.041}$ \\
        B275 SE & 1.2$\pm$0.15 & <60 & -- & -- & $0.65^{0.25} _{-0.25}$ \\
        B331 & <2.9 & <51 & $3.97^{0.37}$ & 10 & 0.02 \\
        B331 NW & 1.9$\pm$0.22 & <51 & -- & -- & $0.65^{0.25} _{-0.25}$ \\
        B331 SW & 3.0$\pm$0.3 & <51 & -- & -- & $0.65^{0.25} _{-0.25}$ \\
        B243 & 1.8$\pm$0.2 & <68 & $2.58^{0.11} _{-0.03}$ & $4.2^{0.4} _{-0.2}$ & $0.98^{0.17} _{-0.21}$ \\
        B243 SW & <0.68 & <68 & -- & -- & $0.65^{0.25} _{-0.25}$ \\
        B268 & <0.81 & -- & $2.66^{0.08} _{-0.04} $& $4.5^{0.3} _{-0.2}$ & $0.74^{0.13} _{-0.12}$\\
    \end{tabular}
    \end{minipage}
    \label{tab:comparison_data}
\end{table}

In the four panels of \Cref{fig:disk_mass_comparison} we show the dust mass in the disk versus stellar mass (upper left), age (upper right), stellar luminosity (lower left), and outer dust radius (lower right). The parameters used for this comparison are summarized in \Cref{tab:comparison_data}. Compared to the other samples, our sample is small. The purpose of this comparison is therefore to see to what extent the higher mass objects in M17 follow previously identified trends.  

In low mass SFRs two main trends have been observed with respect to dust masses among Class II disk populations \citep{lada1987}. The first is an increase with stellar mass, and the second is a decrease with stellar or region age \citep[e.g.,][]{pascucci2016,ansdell2017,barenfeld2016}. The stellar mass trend can be seen in the Lupus and Upper Sco data points in the upper left panel of \Cref{fig:disk_mass_comparison}, the age trend is seen when comparing between those regions globally (upper right panel). The Herbig sample of \cite{stapper2022} follows the positive correlation with stellar mass, but not the negative correlation with older age. The authors explain this by suggesting that the effects of mass and age can be degenerate. As also found in previous studies \citep[e.g.,][]{vandermarel2021}, stars with higher mass do not only start out with a more massive disk, but also appear to retain this mass for a longer time. \cite{vandermarel2021} find that the existence of dust traps, caused by structures such as an inner gap or ring-shaped clearing, is necessary to retain an observable mm-dust reservoir for a longer time. Both \cite{stapper2022} and \cite{vandermarel2023} propose that such structures may form more readily in disks with higher mass.

\begin{table}
    \caption{Mean disk dust masses and mean stellar masses for the samples from five star-forming regions plotted in \Cref{fig:disk_mass_comparison} and discussed in the text.}
   \begin{minipage}{\hsize}
\centering
\renewcommand{\arraystretch}{1.4}
\setlength{\tabcolsep}{3pt}
    \begin{tabular}{l|ccccc}
    \hline
\hline
        Region/Sample & $\rm <M_{dust}>$ & $\sigma$\footnote{Standard deviation of the averaged values.} & $\rm <M_{*}>$ & $\sigma$$^a$ & age \\
        & $ \rm (M_{\oplus})$ & $ \rm (M_{\oplus})$ & $\rm (M_{\odot})$ &  $\rm (M_{\odot})$ & (Myr) \\    
        \midrule
        M17 & 1.7 & 0.82 & 6.5 & 2.3 & 0.6$\pm$0.3 \\
        Upper Sco & 3.9 & 15 & 0.33 & 0.29 & 5-11  \\
        ONC\footnote{Orion Nebula Cluster sample from \cite{eisner2018}, not plotted in \Cref{fig:disk_mass_comparison}.} & 8.2 & 14 & 0.78 & 2.3 & $\sim$1 \\
        Lupus & 17 & 28 & 0.45 & 0.50 & 1-3 \\
         Herbig sample\footnote{The main sample from \cite{stapper2022} (blue marks in \Cref{fig:disk_mass_comparison}), excluding the (higher mass) exceptions (gray marks in \Cref{fig:disk_mass_comparison}; see also \Cref{sec:higher_mass_herbigs}).} & 40 & 48 & 2.0 & 0.38 & 7$\pm$3\\
    \end{tabular}
    \end{minipage}
    \label{tab:mean_values_for_comparison}
\end{table}

There is a large spread in dust mass in the shown samples and as such the dust masses of the M17 objects are not outliers. However, the M17 sample studied here is both the most massive and the youngest sample in question, yet of all the samples has the lowest mean disk mass, contrary to the described trends above. The mean values are listed in \Cref{tab:mean_values_for_comparison} and shown in \Cref{fig:disk_mass_comparison}. Also, the models suggest a mm-dust gap, yet for the M17 objects this does not come with an extended, high-mass mm-dust disk. On the other hand, in the intermediate- to high-mass range the evolutionary timescale of the star may approach or become small with respect to disk evolution timescales of low mass stars. Higher mass stars may sooner deplete their disks due to short formation timescales, high accretion rates, strong UV illumination, and outflows from the star and/or disk. To allow for these effects in the age comparison across the stellar mass ranges, in \Cref{fig:normalized_age_vs_dm} we plot the age of the objects "normalized" by the PMS lifetime corresponding to their stellar mass. The dust mass in this plot is "normalized" by the stellar mass to better isolate the effects of age. (Almost) all plotted objects appear indeed on their PMS (age/ZAMS-age$<1$). As may be expected, the M17 objects are further in their PMS evolution than the low mass star populations, where the PMS lifetime is (very) long ($10^7$ to almost $10^9$\,Myrs) compared to disk evolution timescales. Compared to the Herbig sample from \cite{stapper2022} the M17 objects are similar, or slightly earlier in their evolution. Though there are some Herbig stars that fall in the same range, with respect to all the samples the stellar-mass-normalized dust masses of the M17 sample are very low.

The age-plot in \Cref{fig:disk_mass_comparison} includes the serendipitous discoveries which, as discussed in \Cref{sec:discussion_serendipitous_discoveries}, are likely lower mass YSOs with disks comparable in mass to the original targets. As argued in that section, considering the density of YSOs in the region, the lack of higher-mass disk discoveries may hint at a disk population with relatively low masses, especially for a SFR as young as M17.

One possible explanation for the findings above is that the formation environment plays a crucial role in the initial disk mass distribution as well as disk evolution and lifetime. This explanation is supported by studies in other massive SFRs. \cite{eisner2018}, for example, study the disk population (92 disks) in the Orion Nebula Cluster and find compact disks ($<60$\,au), with low masses and with only a weak correlation between stellar mass and disk mass compared to low-density SFRs (see \Cref{tab:mean_values_for_comparison} for mean disk and stellar masses of their sample). They suggest that the disks are affected by photo-ionization and, possibly, stellar encounters. Similar population studies in Orion suggest that even intermediate far-UV radiation fields from A0 and B stars have a significant impact on the evolution of protoplanetary disks \citep{vanterwisga2020,vanterwisga2023}. Investigating mass and accretion in the disk population of the Lagoon Nebula (M8), \cite{venuti2024} also find evidence for faster disk evolution in the central regions with respect to the outskirts of this massive SFR. 

Finally, the upper limits on the disk radii are consistent with the observed trend between disk mass and radius in \cite{stapper2022}. Given the discussion above it would appear likely that the detections comprise optically thick, compact disks ($\lessapprox 30$\,au), that are able to shield themselves to survive the high UV-radiation environment. The mass estimate even in this case remains rather low. Next to external evaporation and encounters in the dense environment after disk formation, this might also be due to processes like competitive accretion limiting the amount of mass the disks could assemble during disk formation \citep[e.g.,][]{bate2018}.

\subsubsection{Herbig stars of higher mass} \label{sec:higher_mass_herbigs}

\begin{table}
    \caption{Parameters for high mass Herbig stars from \cite{stapper2022}.}
   \begin{minipage}{\hsize}
\centering
\renewcommand{\arraystretch}{1.4}
\setlength{\tabcolsep}{3pt}
    \begin{tabular}{l|ccccc}
    \hline
\hline
        Object & $\rm M_{dust}$ & $\rm R_{\rm dust}$ &  $\log L/L_{\odot}$ & $\rm M_{*}$ & age \\
        & $ \rm (M_{\oplus})$ & au & & $\rm (M_{\odot})$ & (Myr) \\    
        \midrule
        HD58647 & 1.0$\pm$0.1 & <126 & $2.44^{0.11}_{-0.09}$  & $3.9^{0.3}_{-0.2}$  & $0.84^{0.12}_{-0.18}$ \\
        MWC297  & 65.7$\pm$9.6 & <253 & $4.41^{0.39}_{-0.50}$  & $14^{6}_{-5}$ & $0.04^{0.07}_{-0.02}$ \\    
        ZCMa    & 7.6$\pm$6.4 & <41   & $2.25^{0.51}_{-0.29}$  & $3.8^{2.0} _{-0.8}$  & $0.80^{0.83}_{-0.59}$ \\
        \hline
    \end{tabular}
    \end{minipage}
    \label{tab:high_mass_herbig_params}
\end{table}

In addition to the influence of the SFR environment on the disks, another (not mutually exclusive) explanation for the low disk masses of the M17 targets could be a faster disk evolution and dissipation in the higher stellar mass ranges. So far, in our comparison we have taken into account the bulk of the Herbig stars in \cite{stapper2022}, which have a mean mass of $\sim$\,2\,\msun\,(the blue marks in \Cref{fig:disk_mass_comparison}, see also \Cref{tab:mean_values_for_comparison}), significantly lower than the masses of the M17 targets. \cite{stapper2022} marked a few objects in their sample as exceptions (grey marks on the plots) because their Herbig nature was ambiguous - three of these due to their high-mass or very young nature. These three "exceptions" (MWC297, ZCMa, and HD58647) are labeled in the plots, because they are similar to the M17 objects in age, mass, and luminosity (lower left panel in \Cref{fig:disk_mass_comparison} and \Cref{tab:high_mass_herbig_params}).

The low number of objects of this nature stresses the dearth of YSOs in the higher mass ranges ($\gtrsim 4$\,\msun) that have been subject to multi-wavelength studies. Because of this, it is not possible at the moment to derive statistical properties, and the comparison remains on an object-to-object basis. The difference between the labeled sources and the M17 targets is that the former are relatively isolated objects while the latter are located in a dense cluster environment. We now discuss each labeled source separately.  

The two objects ZCMa and HD58647 are similar in mass ($\sim 3.8$ \Msun) and both have outflows associated with accretion. ZCMa has a FU Orionis variable companion and an asymmetric outflow \citep{baines2006}.  HD58647 is observed to have a disk wind from an accretion disk \citep{kurosawa2016} and its dust mass is similar to the M17 objects. This serves as an example that low (observable) dust mass and ongoing accretion can both be present.  

MWC297 is a high mass star ($\sim$\,14.5\,\Msun) in the Herbig catalogue of \cite{vioque2018}, but was earlier classified as a zero-age-main-sequence star of $\sim$\,10\,\msun\,by \cite{drew1997}. \cite{manoj2007} detect a remnant formation disk ($<80$\,au) with dust mass similar to that reported by \cite{stapper2022} ($\sim$\,100\,\Mearth). They note the possibility of a disk wind or stellar wind based on a VLA spectral index of $\sim$\,0.6. Interestingly, while a hot ($\sim$\,1500\,K) gas component from the inner disk ($\sim$\,12\,au) was detected in CO ro-vibrational fundamental and overtone emission, no cold gas was detected in CO lines at mm-wavelengths \citep{sandell2023,manoj2007}. 

MWC297 is very similar to B331 in stellar mass, age, and luminosity. Moreover both have CO overtone bandhead emission in the NIR, both lack cold CO gas, and they have a similar VLA spectral index. The only significant difference is that the dust disk of MWC297 is almost two orders of magnitude higher in mass. Perhaps this points to 
a role for the environment in the disk evolution of high-mass YSOs.

\subsection{Serendipitous discovery of sources} \label{sec:discussion_serendipitous_discoveries}

In the four fields surrounding the original targets four serendipitous detections were made in both ALMA Bands. All these radio sources have NIR counterparts. We collected
NIR and MIR photometry from VizieR Catalogues \footnote{\url{https://vizier.cds.unistra.fr/viz-bin/VizieR}}. The flux points are listed in \Cref{tab:photometry_new_dets}and plotted in \Cref{fig:sed_fits}. The source positions and projected distances from the closest original target are listed in \Cref{tab:source_positions_separations}.

We estimated the expected number of extra-galactic background sources as a function of field-size by adopting Equation A10 from \cite{anglada1998} with the ALMA 12m primary beam $\theta_{\rm A} \approx 20.6'' \times (300/\nu) \rm \,GHz$. This yields (similar to Equation A11 from \cite{anglada1998})
\begin{align*}
    <N> &= 0.00196 \, \left\{ 1 - \exp \left[-0.00123 \left(\frac{\theta_{\rm F}}{\rm arcsec} \right)^2 \left(\frac{\nu}{\rm 300\, GHz} \right)^2 \right] \right\} \\ &\left(\frac{\nu}{\rm 300\, GHz} \right)^{-2} \left(\frac{\rm S_0}{\rm mJy} \right)^{-0.75} \left(\frac{\nu}{\rm 5\, GHz} \right)^{-0.525} \, ,
\end{align*}
with $\rm S_0=0.165 \rm \,mJy$ the minimum detectable flux density at the field center, taken to be five times the sensitivity in Band 6; $\nu=225$\,GHz the Band 6 frequency; and $\theta_{\rm F}=14''$ the field diameter. We find the expected number of extra-galactic sources in each field to be negligible, $<N>\approx 2\times10^{-4}$. 

We now discuss known characteristics per source. 

The two sources B275 SE and B243 SW are listed in the Massive Young star-forming Complex Study in Infrared and X-rays \citep[MYStIX;][]{broos2013} catalog as an X-ray (Chandra) source with NIR counterpart, and classified as "young star in a massive SFR". Both objects also feature in a study by \cite{yang2022a}, who use machine learning techniques to assign probabilities (P) to the classification of X-ray sources based on multi-wavelength features (X-ray, NIR-MIR photometry and colors). They classify B275 SE as low-mass X-ray binary (LMBX) with $\rm P=0.4$ or as YSO with $\rm P=0.3$. Considering the evolved nature of LMBX's and the young age of M17, the first classification seems rather unlikely. B243 SW is classified as YSO with $\rm P=0.75$. This object is also labeled as a variable source in {\it Gaia} DR3 \cite{gaiacollaboration2023}.

Based on these findings, both B275 SE and B243 SW are likely (low mass) YSOs. The spectral indices ($\alpha = 1.4\pm0.55$ and $\alpha = 0.6\pm0.8$ respectively) indicate a contribution from free-free emission, for B275 SE probably next to dust emission from a circumstellar disk. In the case of B243 SW, the presence of both variability and X-rays may indicate that the free-free emission is related to an outflow driven by ongoing accretion \citep[see, e.g.,][]{rota2024}. However, even without correcting for the contribution of free-free flux, the derived dust mass for this detection is rather low (see \Cref{tab:fluxes_indeces_masses}). 

B331 NW and B331 SW have spectral indices well above 2, consistent with (partially) optically thin dust emission from a circumstellar disk. B331 NW ($\alpha = 2.3\pm0.6$) has a relatively low projected separation from B331 ($1.4''$, $\sim$\,2300\,au) and is therefore not resolved in most photometric surveys. In the reported data from Pan-STARRS1 \citep{chambers2016} and MYStIX crowded fields \citep{king2013} it is barely resolved, and the measured fluxes are likely contaminated by B331. It is clearly observed as a distinct source in the previously mentioned acquisition images from LBT (\Cref{fig:acq_B331}). B331 SW ($\alpha = 2.4\pm0.5$) is the faintest detection in the NIR with only one survey \citep{king2013} reporting fluxes, yet it is the second (after B331) brightest detection with ALMA. Perhaps this object is of lower mass, partly obscured by its disk, or simply more embedded. 

All in all, it is likely that all four serendipitous detections relate to (lower-mass?) YSOs and that at least two have a dust disk. None of these objects is resolved, and the measured fluxes are similar to those of the original targets, leading to again similar values for (upper limits on) the dust mass and radius. For the reported masses in \Cref{tab:fluxes_indeces_masses} a temperature of $T=$\,150\,K was used. Assuming lower dust temperatures for these sources, e.g., $T_{\rm dust}=$\,20\,K, would result in dust mass estimates about one order of magnitude higher (i.e., a few tens of \Mearth), closer to the averages found in low-mass SFRs (\Cref{fig:disk_mass_comparison}). Yet for the two objects with free-free emission this would still represent an upper limit. 

The overall picture from these detections, under the assumption that they are YSOs, is that their disks are small and on the lower mass end for an age of $\sim 0.6~\rm Myrs$ (see upper right panel \Cref{fig:disk_mass_comparison}). While the discoveries testify to the source density in the region, it is remarkable that brighter disks that would have been detected if present in these fields, are lacking. The number of sources is too small to be conclusive, but it is not unlikely that we are "catching" the brightest (i.e., most massive) disks and that deeper observations would yield a yet fainter population. As suggested before for the original targets it is possible that the disks are compact, optically thick, and more massive than derived under an optically thin assumption. That would support the notion that in a dense, high-UV environment like M17 perhaps this is the kind of disk that survives, when disks survive at all. 

\section{Summary and conclusions} \label{sec:conclusions}

In this paper, we present ALMA mm-continuum detections of four intermediate- to high-mass PMS stars in the massive star forming H~{\sc ii} region M17 (distance 1.7\,kpc) and report on the serendipitous discovery of four additional sources, likely low-mass YSOs. All detected sources are unresolved at a resolution of $\sim$\,$0.07''$, constraining the outer radius of the emission to maximally $\sim$\,60\,au. We use the spectral index between ALMA Bands 6 and 7, and (upper limits on) the VLA cm-flux to determine the origin of the mm-emission, that is, free-free radiation or thermal dust emission. We derive (upper limits on the) dust masses from the ALMA continuum flux and an upper limit on the gas mass from the non-detection of the rotational $^{12}\rm CO$ ($J\,=\,2-1$) line. We combine the ALMA data with near- to mid-infrared photometry and spectroscopy and apply different models to estimate the total disk masses of the target sources. In the following we summarize our results:

\begin{itemize}
    \item The disks around the four target sources as well as the four serendipitous discoveries contain a dust mass of (at most) a few \Mearth. 
     \item Estimates of (the upper limit on) the total gas mass in the disks around the intermediate- to high-mass PMS stars vary between $\sim 2 \times 10^{-5}~M_{\odot}$, and, when assuming an optically thick compact disk, $\sim 6 \times 10^{-3}~M_{\odot}$.
    \item Spectral indices suggest that the fluxes from two target sources and two serendipitous sources have a significant contribution from free-free emission from ionized material. In the higher-mass sources (B331 and B268) this may indicate a small-scale \Hii-region, the base of a radio jet, or a photo-evaporative disk wind. In the (likely) low mass sources (B275~SE and B243~SW) the combination with X-ray emission may indicate an accretion driven outflow.  
    \item Our modeling suggests that the inner disks of (some of) the targets are dust depleted, reminiscent of gap- or ring-like structures, despite the presence of hot and dense gaseous disks, as evidenced by CO bandhead emission from these objects.
    \item Comparison to disk population studies in low-mass SFRs and among Herbig stars indicates that our sample is both the most massive and the youngest. However, contrary to commonly observed trends, it has the lowest mean disk dust mass.  
\end{itemize}
All in all, we conclude that the studied targets are surrounded by low-mass, compact disks that likely do not offer a significant contribution anymore to either the final stellar mass or the formation of a planetary system. Along with the four serendipitous discoveries, our findings offer tentative evidence of the influence of the massive star formation environment on disk formation, lifetime, and evolution. The detections presented hint at a rich YSO population in M17, with disks detectable by ALMA. This, along with the thousands of known IR excess sources in M17 \citep{lada1991,jiang2002}, offers the perspective of disk population studies that not only include a significant number of intermediate- to high-mass targets, but also allows the study of them along side low-mass YSOs in a massive SFR.

\begin{acknowledgements}
The authors express their gratitude to the anonymous referee for their helpful comments and insights toward improving this paper. \\ JP acknowledges support from NWO-FAPESP grant 629.004.001 (PI L. Kaper). This paper makes use of the following ALMA data: 2019.1.00910.S, 2018.1.01091.S. ALMA is a partnership of ESO (representing its member states), NSF (USA) and NINS (Japan), together with NRC (Canada), MOST and ASIAA (Taiwan), and KASI (Republic of Korea), in cooperation with the Republic of Chile. The Joint ALMA Observatory is operated by ESO, AUI/NRAO, and NAOJ. \\
This publication makes use of data products from the Two Micron All Sky Survey, which is a joint project of the University of Massachusetts and the Infrared Processing and Analysis Center/California Institute of Technology, funded by the National Aeronautics and Space Administration and the National Science Foundation. \\
This work has made use of data from the European Space Agency (ESA) mission {\it Gaia} (\url{https://www.cosmos.esa.int/gaia}), processed by the {\it Gaia} Data Processing and Analysis Consortium (DPAC, \url{https://www.cosmos.esa.int/web/gaia/dpac/consortium}). Funding for the DPAC
has been provided by national institutions, in particular the institutions
participating in the {\it Gaia} Multilateral Agreement. \\
This research has made use of the VizieR catalog access tool, CDS, Strasbourg, France (DOI : 10.26093/cds/vizier). \\
CHR acknowledges the support of the Deutsche Forschungsgemeinschaft (DFG, German Research Foundation) Research Unit ``Transition discs'' - 325594231 and the support by the Excellence Cluster ORIGINS which is funded by the DFG under Germany's Excellence Strategy - EXC-2094 - 390783311. CHR is grateful for support from the Max Planck Society.
\end{acknowledgements}

\bibliography{ALMA_paper}

\begin{appendix}
\section{Observational details} \label{sec:obs_details}
\begin{table}[h!]
\footnotesize
\centering
\caption{Observational details per target.}        
\begin{minipage}{0.7\hsize}
\centering
\renewcommand{\arraystretch}{1.4}
\setlength{\tabcolsep}{3pt}
\begin{tabular}{l c c c c c c c c}
ALMA source name & RA & Dec & Band & MRS & FOV & Int. Time & L5 BL & L80 BL \\
\hline
Cl\_star\_NGC\_6618\_B243 & 18:20:26.560 & -16:10:03.332 & 6 & 0.804 & 25.830 & 1415.232 & 373.619 & 4517.446 \\
Cl\_star\_NGC6618\_B331 & 18:20:21.634 & -16:11:17.980 & 6 & 0.814 & 25.831 & 1415.199 & 371.527 & 4512.133 \\
Cl\_star\_NGC6618\_B275 & 18:20:25.027 & -16:10:26.134 & 6 & 0.819 & 25.830 & 1415.232 & 369.534 & 4495.101 \\
Cl\_star\_NGC6618\_B268 & 18:20:25.287 & -16:10:18.517 & 6 & 0.808 & 25.830 & 1415.232 & 372.249 & 4507.622 \\
Cl\_star\_NGC\_6618\_B243 & 18:20:26.560 & -16:10:03.332 & 7 & 0.755 & 17.055 & 1233.792 & 272.415 & 2877.893 \\
Cl\_star\_NGC6618\_B331 & 18:20:21.634 & -16:11:17.980 & 7 & 0.752 & 17.055 & 1161.205 & 273.165 & 2886.794 \\
Cl\_star\_NGC6618\_B275 & 18:20:25.027 & -16:10:26.134 & 7 & 0.751 & 17.055 & 1124.928 & 273.863 & 2883.192 \\
Cl\_star\_NGC6618\_B268 & 18:20:25.287 & -16:10:18.517 & 7 & 0.745 & 17.055 & 1124.928 & 275.502 & 2892.298 \\
\end{tabular}
\end{minipage}
\label{tab:observational_details}
\normalsize
\end{table}

\section{Full fitting results} \label{sec:full_fitting_results}
In Tables \ref{tab:gaussian_fitting_results_1} and \ref{tab:gaussian_fitting_results_2} we provide the detailed results of the 2D Gaussian fitting procedure described in \Cref{sec:detections_fluxes}. Where the values for the deconvolved major/minor axes are missing, the CASA fitting task could not deconvolve the source from the beam, due to insufficient resolution. The cases where the deconvolved axes are provided, the resulting values are significantly smaller than the corresponding beam axis and likely not representative of true source sizes.

\begin{sidewaystable*}[hp]
    \centering
    \caption{2D Gaussian CASA fitting results Band 6.}
    \renewcommand{\arraystretch}{1.35}
    \begin{tabular}{@{}lcccccccc@{}}
        \toprule
        Parameter & B275 & B275 SE & B331 & B331 NW & B331 SW & B243 & B243 SW & B268 \\ \midrule
        RMS input (Jy) & 7.78 $\times 10^{-5}$ & 9.43 $\times 10^{-5}$ & 4.31 $\times 10^{-4}$ & 7.27 $\times 10^{-5}$ & 1.66 $\times 10^{-4}$ & 6.87 $\times 10^{-5}$ & 8.14 $\times 10^{-5}$ & 6.18 $\times 10^{-5}$ \\
        RMS residual (Jy) & 1.63 $\times 10^{-5}$ & 2.07 $\times 10^{-5}$ & 2.93 $\times 10^{-5}$ & 2.23 $\times 10^{-5}$ & 3.39 $\times 10^{-5}$ & 1.91 $\times 10^{-5}$ & 1.47 $\times 10^{-5}$ & 1.26 $\times 10^{-5}$ \\
        Zero Level Offset ($\mu$Jy/beam) & -7.74$\pm$1.05 & -28.6$\pm$4.78 & 22.2$\pm$2.44 & -9.86$\pm$1.52 & -17.6$\pm$9.13  & 8.76$\pm $2.16 & 12.9 $\pm$1.76  & -15.6$\pm$2.32  \\
        Integrated Flux ($\mu$Jy) & 434$\pm$40 & 400$\pm$51 & 1698$\pm$57 & 276$\pm$38 & 460$\pm$64 & 252$\pm$47 & 218$\pm$27 & 242$\pm$35 \\
        Peak Flux ($\mu$Jy/beam) & 279$\pm$16 & 250$\pm$21 & 1485$\pm$29 & 280$\pm$21 & 410$\pm$33 & 158$\pm$18 & 201$\pm$14 & 138$\pm$13 \\
        Major Axis (convolved) (mas) & 88.0$\pm$6.7 & 90.7$\pm$9.8 & 59.76$\pm$1.46 & 53.7$\pm$5.1 & 60.4$\pm$6.2 & 94.2$\pm$15.1 & 71.9$\pm$6.8 & 205$\pm$23 \\
        Minor Axis (convolved) (mas) & 45.0$\pm$2.0 & 44.9$\pm$2.8 & 36.91$\pm$0.57 & 35.4$\pm$2.2 & 35.8$\pm$2.3 & 42.2$\pm$3.6 & 37.6$\pm$1.9 & 136$\pm$11 \\
        Position Angle (convolved) (deg) & 98.8$\pm$2.5 & 111.4$\pm$3.3 & 106.9$\pm$1.3 & 84.7$\pm$5.8 & 114.8$\pm$4.7 & 113.2$\pm$3.9 & 98.1$\pm$2.9 & 64.7$\pm$8.0 \\
        Beam Major Axis (arcsec) & 0.07 & 0.07 & 0.06 & 0.06 & 0.06 & 0.06 & 0.06 & 0.16 \\
        Beam Minor Axis (arcsec) & 0.04 & 0.04 & 0.03 & 0.03 & 0.03 & 0.04 & 0.04 & 0.10 \\
        Beam Position Angle (deg) & -76.47 & -76.47 & -74.48 & -74.48 & -74.48 & -76.72 & -76.72 & -72.46 \\
        Major Axis (deconvolved) (mas) & 58.2$\pm$11.0 & 63$\pm$16 & 19.7$\pm$5.4 & -- & -- & 69$\pm$23 & -- & -- \\
        Minor Axis (deconvolved) (mas) & 22.8$\pm$6.4 & 21$\pm$14 & 13.6$\pm$4.5 & -- & -- & 13$\pm$11 & -- & -- \\
        Position Angle (deconvolved) (deg) & 94$\pm$10 & 118$\pm$15 & 121$\pm$70 & -- & -- & 119$\pm$16 & -- & -- \\
        \bottomrule
    \end{tabular}
    \label{tab:gaussian_fitting_results_1}
\vspace{0.55cm}
    \centering
    \caption{2D Gaussian CASA fitting results Band 7.}
    \begin{tabular}{@{}lcccccccc@{}}
        \toprule
        Parameter & B275 & B275 SE & B331 & B331 NW & B331 SW & B331 NE & B243 & B243 SW \\ \midrule
        RMS input (Jy) & 1.82 $\times 10^{-4}$& 2.23 $\times 10^{-4}$ & 6.15$\times 10^{-4}$ & 2.31$\times 10^{-4}$ & 5.91 $\times 10^{-4}$ & 3.31 $\times 10^{-4}$& 1.65$\times 10^{-4}$ & 1.98$\times 10^{-4}$  \\
        RMS residual (Jy) & 4.51$\times 10^{-5}$ & 4.92$\times 10^{-5}$ & 4.86$\times 10^{-5}$ & 4.28$\times 10^{-5}$ & 3.65$\times 10^{-5}$ & 5.92$\times 10^{-5}$ & 3.28$\times 10^{-5}$ & 3.65$\times 10^{-5}$  \\
        Zero Level Offset ($\mu$Jy/beam) & 22.2 $\pm$ 2.95 & 62.0 $\pm$ 6.98 & 34.9 $\pm$ 3.86 & 3.13 $\pm$ 3.95 & 111 $\pm$ 10.4 & 166 $\pm$ 9.48 & 36.2 $\pm$ 2.19 & 10.9 $\pm$ 5.47  \\
        Integrated Flux ($\mu$Jy) & 704 $\pm$ 92 & 374 $\pm$ 72 & 2310 $\pm$ 100 & 806 $\pm$ 80 & 1122 $\pm$ 63 & 601 $\pm$ 133 & 356 $\pm$ 44 & 249 $\pm$ 60   \\
        Peak Flux ($\mu$Jy/beam) & 573 $\pm$ 45 & 448 $\pm$ 45 & 1848 $\pm$ 49 & 721 $\pm$ 40 & 1116 $\pm$ 34 & 439 $\pm$ 60 & 476 $\pm$ 28 & 257 $\pm$ 31   \\
        Major Axis (convolved) (mas) & 82.7 $\pm$ 8.2 & 64.0 $\pm$ 8.1 & 68.18 $\pm$ 2.23 & 71.5 $\pm$ 5.5 & 64.25 $\pm$ 2.64 & 71.2 $\pm$ 12.0 & 68.0 $\pm$ 5.7 & 90.9 $\pm$ 17.6  \\
        Minor Axis (convolved) (mas) & 47.9 $\pm$ 2.9 & 42.2 $\pm$ 3.4 & 42.01 $\pm$ 0.89 & 35.9 $\pm$ 1.4 & 35.88 $\pm$ 0.82 & 44.1 $\pm$ 4.9 & 36.6 $\pm$ 1.5 & 35.4 $\pm$ 2.6  \\
        Position Angle (convolved) (deg) & 122.6 $\pm$ 4.3 & 113.2 $\pm$ 7.6 & 120.5 $\pm$ 1.7 & 124.9 $\pm$ 2.2 & 119.0 $\pm$ 1.5 & 120.3 $\pm$ 9.1 & 123.3 $\pm$ 2.5 & 85.0 $\pm$ 2.8  \\
        Beam Major Axis (arcsec) & 0.07 & 0.07 & 0.06 & 0.06 & 0.06 & 0.06 & 0.08 & 0.08  \\
        Beam Minor Axis (arcsec) & 0.04 & 0.04 & 0.04 & 0.04 & 0.04 & 0.04 & 0.04 & 0.04   \\
        Beam Position Angle (deg) & -63.85 & -63.85 & -60.23 & -60.23 & -60.23 & -60.23 & -60.66 & -60.66  \\
        Major Axis (deconvolved) (mas) & 40 $\pm$ 16 & -- & 25.2 $\pm$ 6.0 & --& -- & 32 $\pm$ 23 & -- & --  \\
        Minor Axis (deconvolved) (mas) & 14 $\pm$ 11 & -- & 21.2 $\pm$ 5.8 & -- & -- & 25 $\pm$ 16 & -- & --  \\
        Position Angle (deconvolved) (deg) & 139 $\pm$ 55 & -- & 132 $\pm$ 89 & -- & -- & 124 $\pm$ 24 & -- & -- \\
        \bottomrule
    \end{tabular}
\label{tab:gaussian_fitting_results_2}
\end{sidewaystable*}

\section{Field images} \label{sec:app_field_images}
\begin{figure*}[hp]
\centering
\includegraphics[width=1\textwidth]{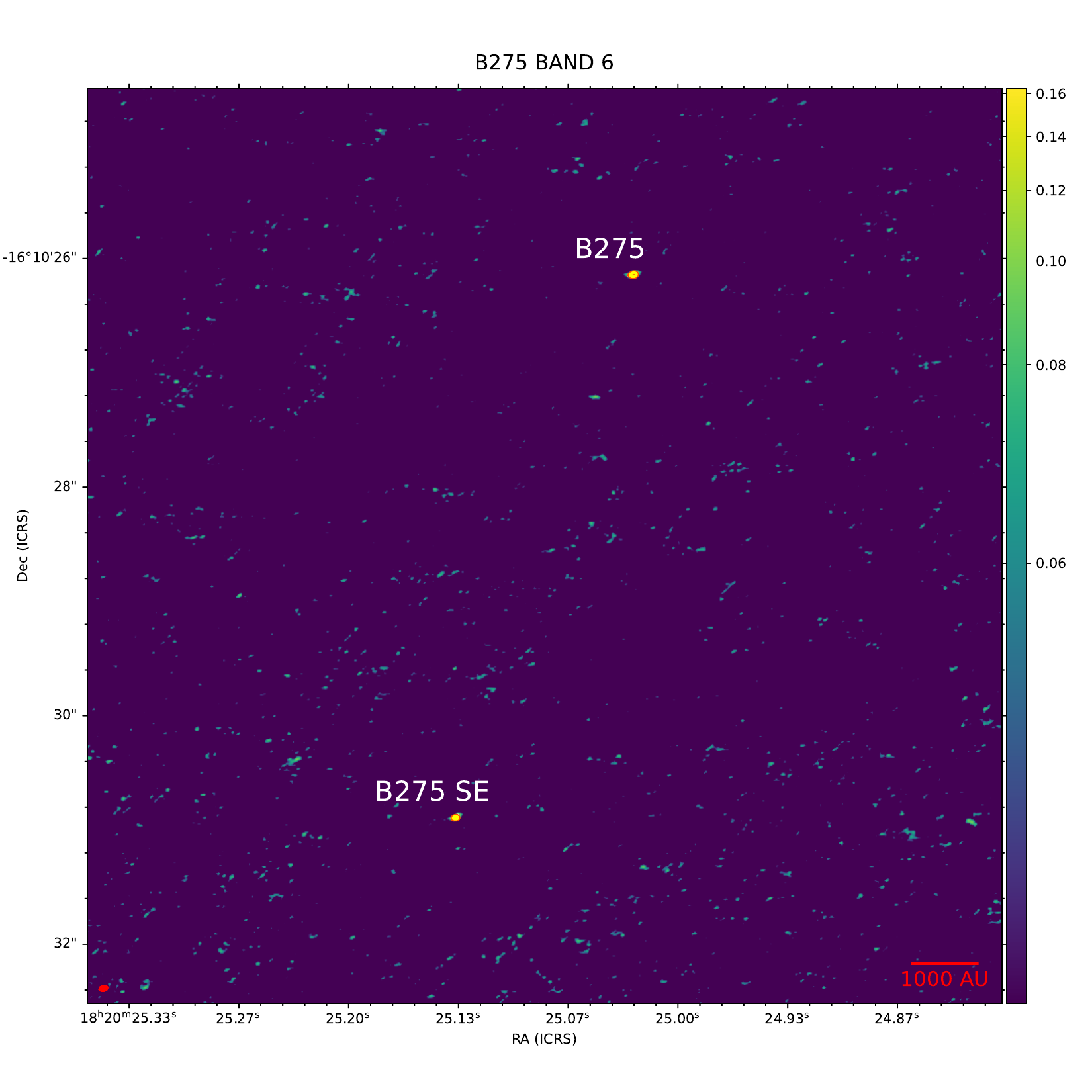}
\caption{Image around target source B275 in ALMA Band 6, showing detections of B275 and B275 SE.}
\label{fig:field_img_B275_6}
\end{figure*}

\begin{figure*}
\centering
 \includegraphics[width=1\textwidth]{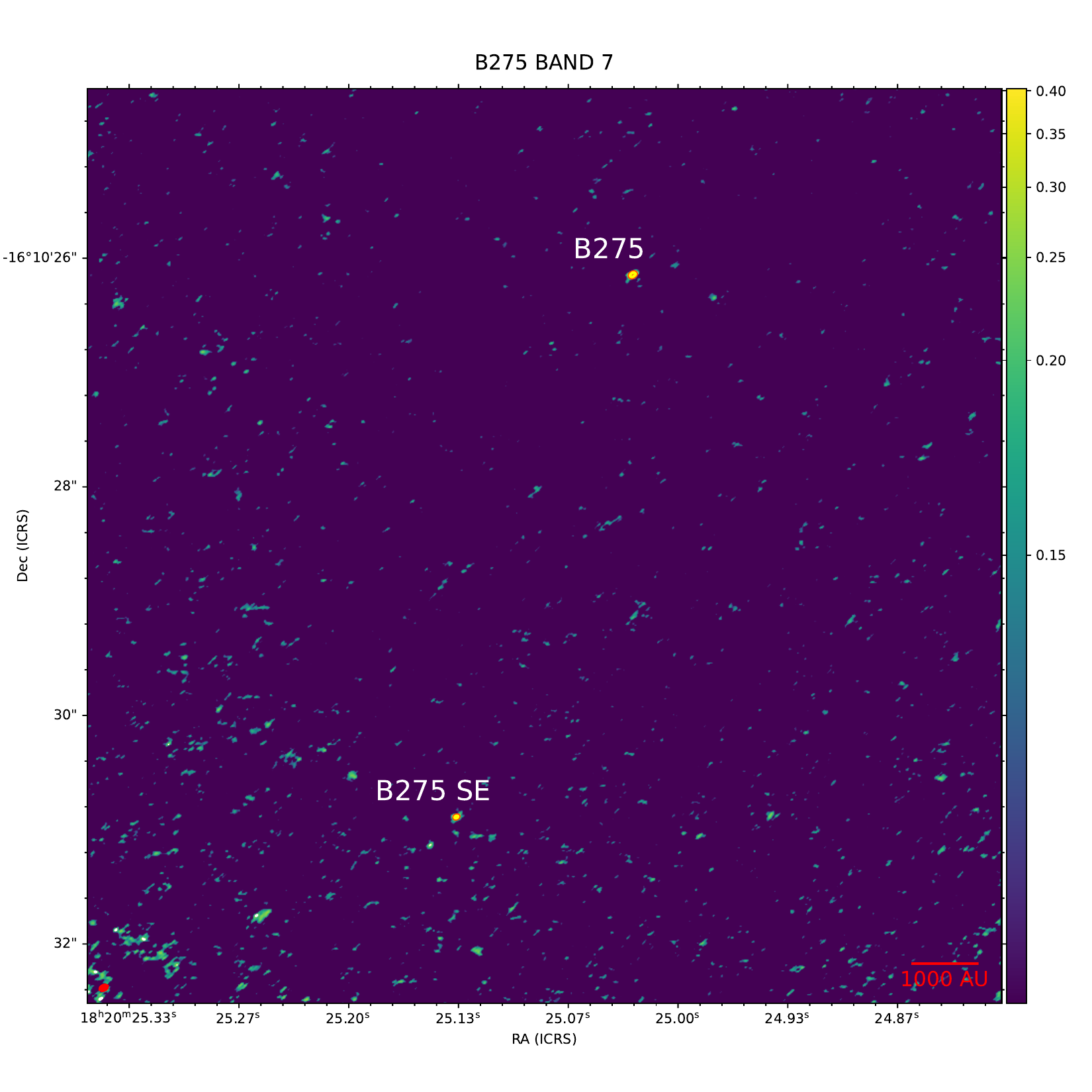} 
\caption{Image around target source B275 in ALMA Band 7, showing detections of B275 and B275 SE.} 
\label{fig:field_img_B275_7}
\end{figure*}

\begin{figure*}
\centering
 \includegraphics[width=1\textwidth]{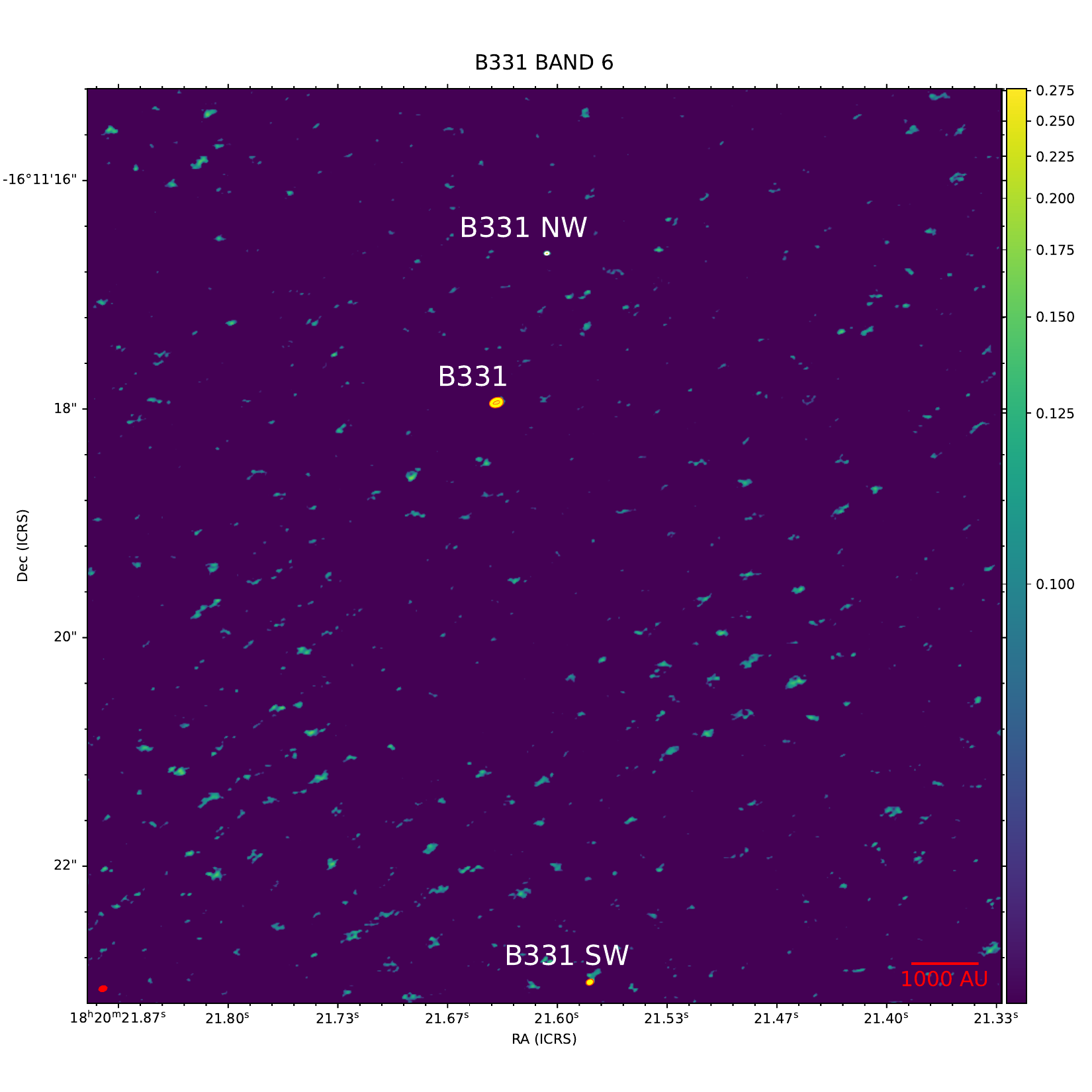} 
\caption{Image around target source B331 in ALMA Band 6, showing detections B331, B331 NW and B331 SW simultaneously.} 
\label{fig:field_img_B331_6_2}
\end{figure*}

\begin{figure*}
\centering
 \includegraphics[width=1\textwidth]{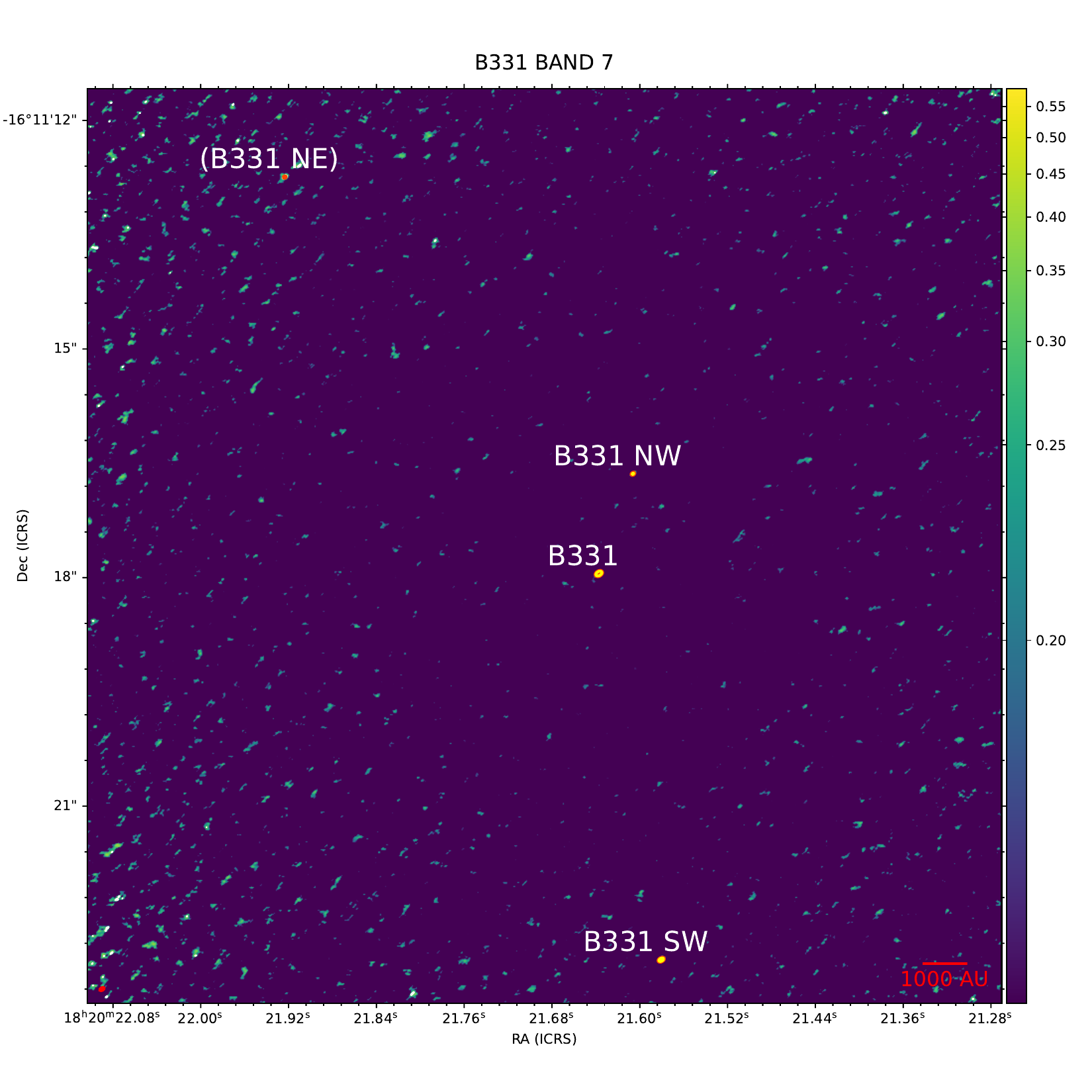} 
\caption{Image around target source B331 in ALMA Band 7, showing detections B331, B331 NW and B331 SW as well as B331 NE, which could also be a noise peak.} 
\label{fig:field_img_B331_7}
\end{figure*}

\begin{figure*}
\centering
\includegraphics[width=1\textwidth]{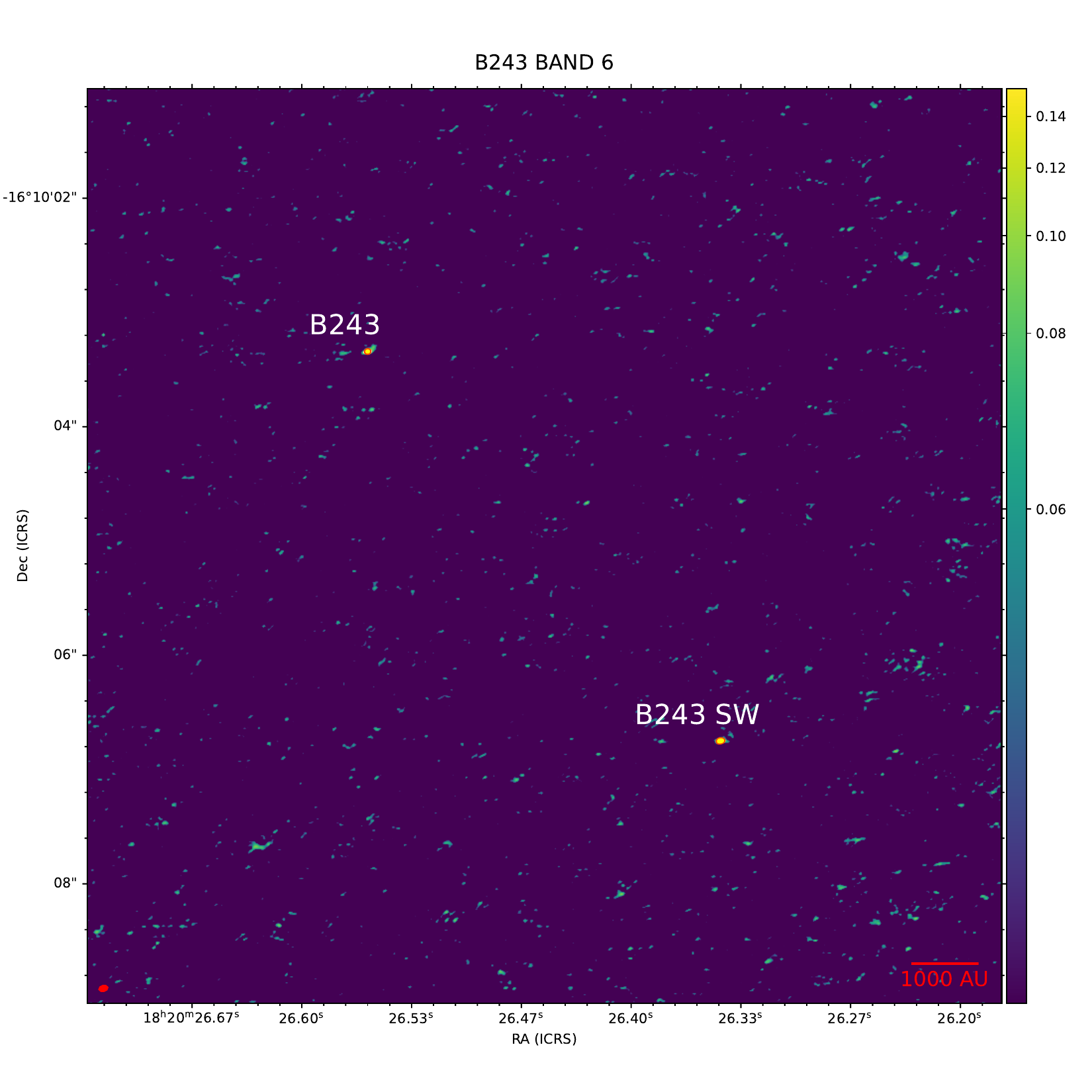}
\caption{Image around target source B243 in ALMA Band 6. B243 and B243 SW are detected on this image.}
\label{fig:field_img_B243_6}
\end{figure*}

\begin{figure*}
\centering
 \includegraphics[width=1\textwidth]{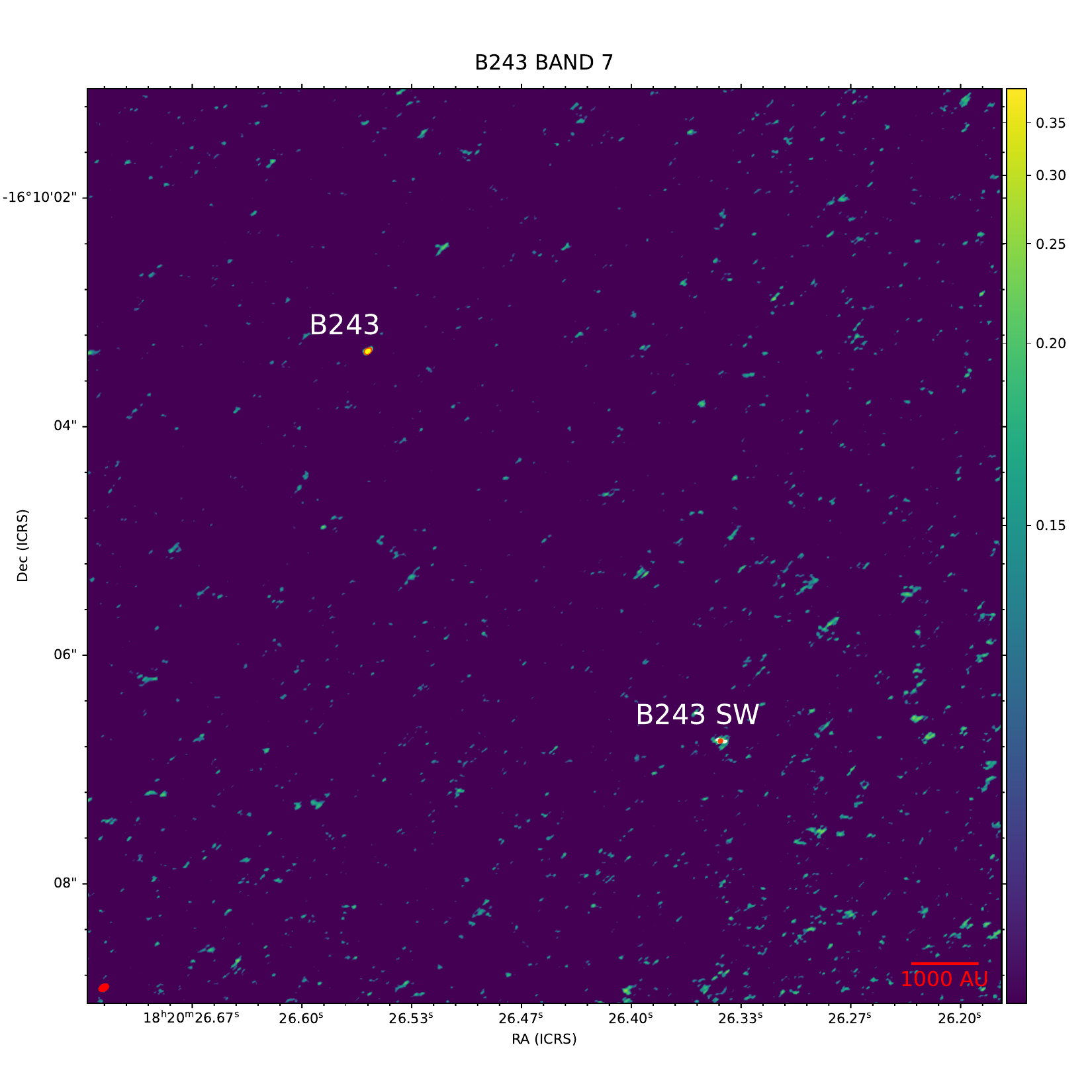} 
\caption{Image around target source B243 in ALMA Band 7. B243 and B243 SW are detected on this image.} 
\label{fig:field_img_B243_7}
\end{figure*}

\begin{figure*}
\centering
\includegraphics[width=1\textwidth]{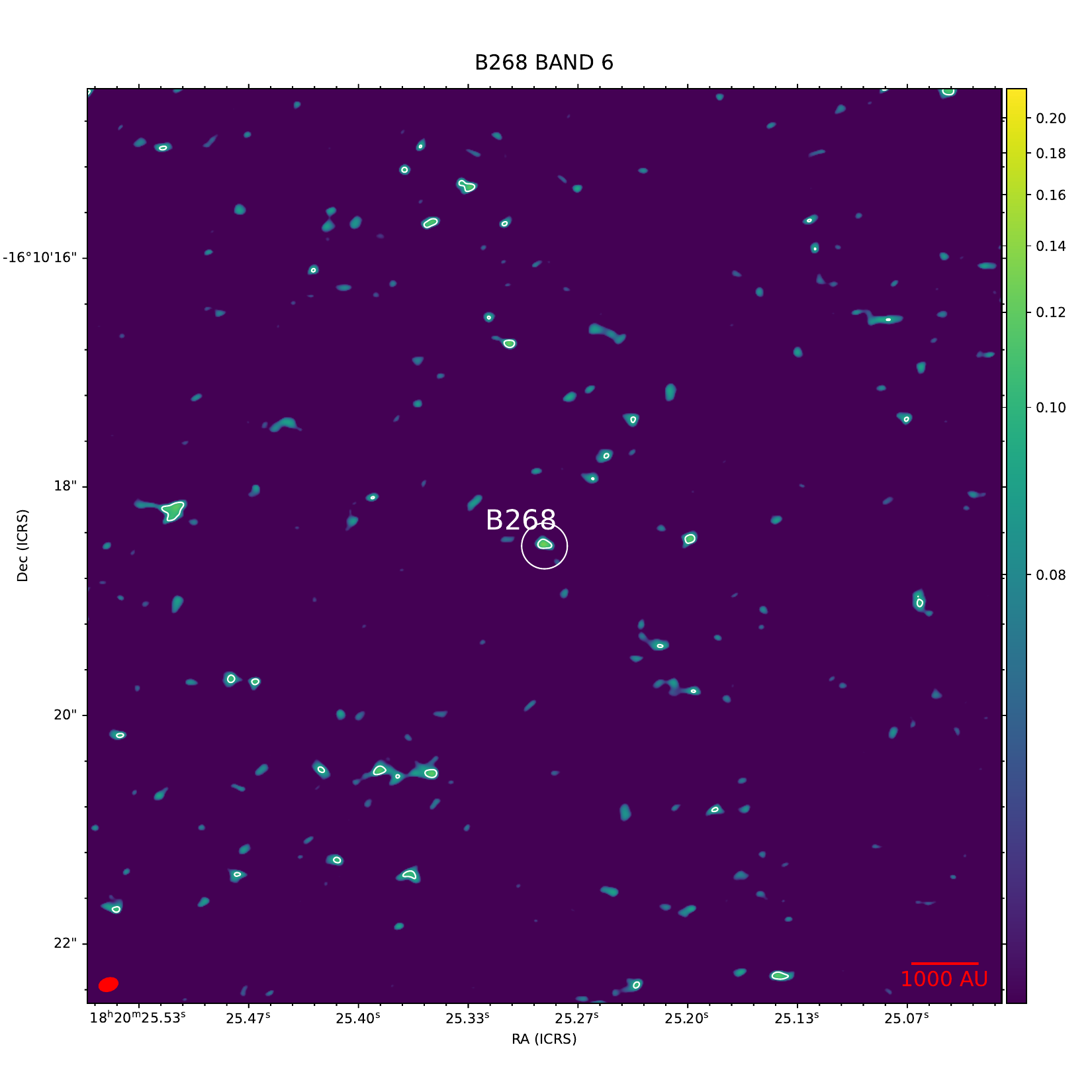}
\caption{Image around target source B268 in ALMA Band 6. B268 is only possibly detected in this image (SNR 3), as the central contour that could otherwise be a noise peak coincides precisely with the coordinates of the NIR source (marked by the white circle).}
\label{fig:field_img_B268_6}
\end{figure*}

\begin{figure*}
\centering
 \includegraphics[width=1\textwidth]{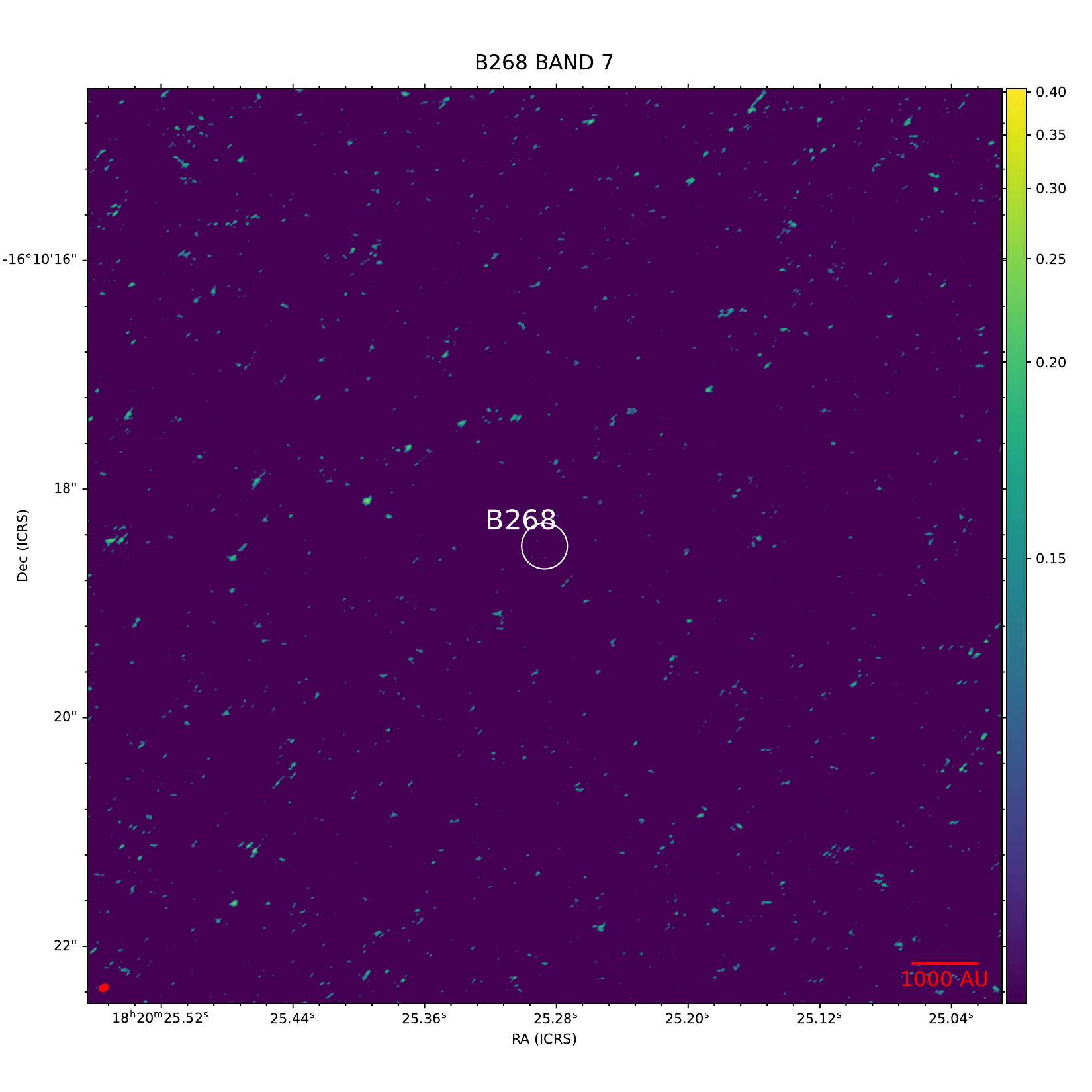} 
\caption{Image around target source B268 in ALMA Band 7. Though the location of B268 is marked for reference, there are no detections in this image.} 
\label{fig:field_img_B268_7}
\end{figure*}

In Figures \ref{fig:field_img_B275_6} to \ref{fig:field_img_B268_7} we show fields from the images created to determine source detections (described in \Cref{sec:alma_observations}). Fields from these images were already presented in \Cref{fig:all_detected_sources} zooming in on each detected source. To clarify the significance of the detections we present a larger field for each target source, in Band 6 and Band 7. Each field shows the area around the detections (or, in case of \Cref{fig:field_img_B268_7}, the lack thereof) with the sources labeled in white, as well as the 5, 7, and 8 $\times ~rms$ contours which are drawn in white, orange-red, and yellow respectively. Only on figure \Cref{fig:field_img_B268_6} the white contours indicate 3 $\times ~rms$. The beam is indicated in the bottom left corner, and a scale bar is shown in the bottom right corner, both in red. The images are scaled such that only emission from $\geq 2\times ~rms$ is non-zero.

\clearpage
\newpage

\section{Acquisition images LBT}
All serendipitous detections have NIR counterparts. Apart from catalogued fluxes, these sources were also detected in the K-band on acquisition images made with the LBT telescope (see \Cref{sec:detections_fluxes}, and Derkink et al. 2024, in prep.). 
\begin{figure}[h]
\centering
\includegraphics[width=1.6\hsize]{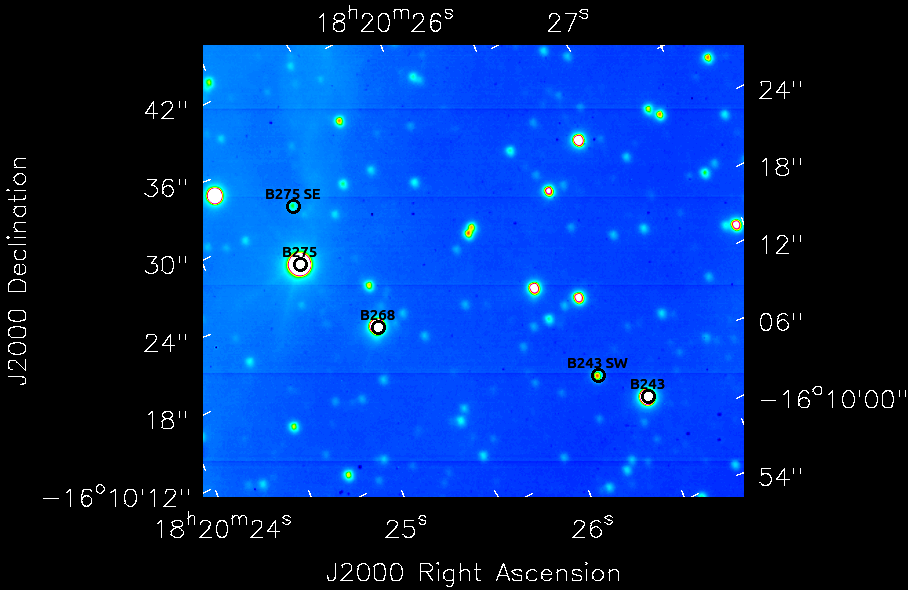}
\caption{Serendipitous detections near B275 and B243.}
\label{fig:acq_B275etc}

 \includegraphics[width=1.6\hsize]{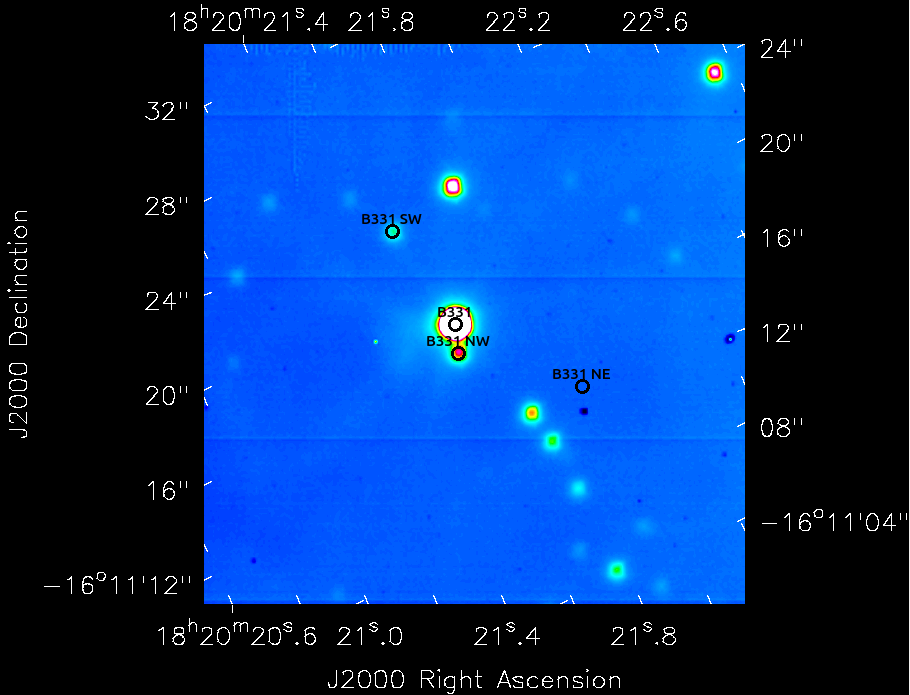} 
\caption{Serendipitous detections near B331.}     
\label{fig:acq_B331}
\end{figure}

\clearpage
\newpage

\section{Photometry}

\begin{table}[h!]
\footnotesize
\centering
\caption{Photometry for the serendipitous detections.}        
\begin{minipage}{1\hsize}
\centering
\renewcommand{\arraystretch}{1.4}
\setlength{\tabcolsep}{3pt}
\begin{tabular}{l c c c c c c c c}
Source   &   freq (GHz)         &  wvl (\micron)        &   flux (mJy)                                    &  filter            & VizieR table         & reference   \\
  \hline                                                                                                
B275 SE  & $ 393.40\times10^3 $ & $ 0.762$   &  $89.3 \pm 70.3\times 10^{-3}$ & GAIA/GAIA3:Grp   & I/355/gaiadr3        &   [1]       \\
B275 SE  & $ 514.90\times10^3 $ & $ 0.582$   &  $18.8 \pm 0.4 \times 10^{-3}$ & GAIA/GAIA3:G     & I/355/gaiadr3        &   [1]       \\
B275 SE  & $ 595.30\times10^3 $ & $ 0.504$   &  $16.4 \pm 6.7 \times 10^{-3}$ & GAIA/GAIA3:Gbp   & I/355/gaiadr3        &   [1]       \\
B275 SE  & $ 135.13\times10^3 $ & $ 2.22 $   &  $4.01 \pm 0.04              $ & UKIRT/WFCAM:K    & J/ApJS/209/28/table2 &   [2]       \\
B275 SE  & $ 183.49\times10^3 $ & $ 1.63 $   &  $2.95 \pm 0.03              $ & UKIRT/WFCAM:H    & J/ApJS/209/28/table2 &   [2]       \\
B275 SE  & $ 240.03\times10^3 $ & $ 1.25 $   &  $1.21 \pm 0.02              $ & UKIRT/WFCAM:J    & J/ApJS/209/28/table2 &   [2]       \\
 \hline                                                                                                                                              
B243 SW  & $ 393.40\times10^3 $ & $ 0.762$   &  $279. \pm 38.\times 10^{-3} $ & GAIA/GAIA3:Grp   & I/355/gaiadr3        &   [1]       \\
B243 SW  & $ 514.90\times10^3 $ & $ 0.582$   &  $43.6 \pm 1.6\times 10^{-3} $ & GAIA/GAIA3:G     & I/355/gaiadr3        &   [1]       \\
B243 SW  & $ 595.30\times10^3 $ & $ 0.504$   &  $37.8 \pm 4.7\times 10^{-3} $ & GAIA/GAIA3:Gbp   & I/355/gaiadr3        &   [1]       \\
B243 SW  & $ 312.40\times10^3 $ & $ 0.960$   &  $1.16 \pm ...               $ & PAN-STARRS/PS1:y & II/349/ps1           &   [3]       \\
B243 SW  & $ 346.50\times10^3 $ & $ 0.865$   &  $83.0 \pm 23.8\times 10^{-3}$ & PAN-STARRS/PS1:z & II/349/ps1           &   [3]       \\
B243 SW  & $ 400.80\times10^3 $ & $ 0.748$   &  $22.7 \pm ...               $ & PAN-STARRS/PS1:i & II/349/ps1           &   [3]       \\
B243 SW  & $ 489.40\times10^3 $ & $ 0.613$   &  $31.8 \pm ...               $ & PAN-STARRS/PS1:r & II/349/ps1           &   [3]       \\
B243 SW  & $ 628.20\times10^3 $ & $ 0.477$   &  $50.7 \pm ...               $ & PAN-STARRS/PS1:g & II/349/ps1           &   [3]       \\
B243 SW  & $ 136.89\times10^3 $ & $ 2.19 $   &  $32.7 \pm 3.8               $ & Johnson:K2MASS   & II/246/out(2MASS)    &   [4]       \\
B243 SW  & $ 136.89\times10^3 $ & $ 2.19 $   &  $6.57 \pm 1.00              $ & Johnson:K2MASS   & II/246/out(2MASS)    &   [4]       \\
B243 SW  & $ 183.92\times10^3 $ & $ 1.63 $   &  $15.4 \pm 1.7               $ & Johnson:H2MASS   & II/246/out(2MASS)    &   [4]       \\
B243 SW  & $ 183.92\times10^3 $ & $ 1.63 $   &  $2.77 \pm 1.08              $ & Johnson:H2MASS   & II/246/out(2MASS)    &   [4]       \\
B243 SW  & $ 239.83\times10^3 $ & $ 1.25 $   &  $8.64\pm ...                $ & Johnson:J2MASS   & II/246/out(2MASS)    &   [4]       \\
B243 SW  & $ 135.13\times10^3 $ & $ 2.22 $   &  $8.18 \pm 0.04              $ & UKIRT/WFCAM:K    & J/ApJS/209/28/table2 &   [2]       \\
B243 SW  & $ 183.49\times10^3 $ & $ 1.63 $   &  $4.13 \pm 0.02              $ & UKIRT/WFCAM:H    & J/ApJS/209/28/table2 &   [2]       \\
B243 SW  & $ 240.03\times10^3 $ & $ 1.25 $   &  $1.23 \pm 0.01              $ & UKIRT/WFCAM:J    & J/ApJS/209/28/table2 &   [2]       \\
B243 SW  & $ 52.311\times10^3 $ & $ 5.73 $   &  $10.4 \pm 2.4               $ & Spitzer/IRAC:5.8 & J/ApJS/209/32/mpcm   &   [5]       \\
B243 SW  & $ 66.724\times10^3 $ & $ 4.49 $   &  $11.3 \pm 1.3               $ & Spitzer/IRAC:4.5 & J/ApJS/209/32/mpcm   &   [5]       \\
B243 SW  & $ 84.449\times10^3 $ & $ 3.55 $   &  $12.9 \pm 1.0               $ & Spitzer/IRAC:3.6 & J/ApJS/209/32/mpcm   &   [5]       \\
   \hline                                                                                                                                           
B331 NW  & $ 312.40\times10^3 $ & $ 0.960 $  &  $4.52 \pm ...               $ & PAN-STARRS/PS1:y & II/349/ps1           &   [3]         \\
B331 NW  & $ 346.50\times10^3 $ & $ 0.865 $  &  $35.9 \pm 2.1\times 10^{-3} $ & PAN-STARRS/PS1:z & II/349/ps1           &   [3]         \\
B331 NW  & $ 135.13\times10^3 $ & $ 2.22  $  &  $54.0 \pm 1.0               $ & UKIRT/WFCAM:K    & J/ApJS/209/28/table2 &   [2]       \\
B331 NW  & $ 183.49\times10^3 $ & $ 1.63  $  &  $23.5 \pm 0.4               $ & UKIRT/WFCAM:H    & J/ApJS/209/28/table2 &   [2]       \\
B331 NW  & $ 240.03\times10^3 $ & $ 1.25  $  &  $24.3 \pm 0.4               $ & UKIRT/WFCAM:J    & J/ApJS/209/28/table2 &   [2]       \\
  \hline                                                                                                                          
B331 SW  & $ 135.13\times10^3 $ & $ 2.22  $  &  $3.67 \pm 0.06              $ & UKIRT/WFCAM:K    & J/ApJS/209/28/table2 &   [2]       \\
B331 SW  & $ 183.49\times10^3 $ & $ 1.63  $  &  $1.79 \pm 0.03              $ & UKIRT/WFCAM:H    & J/ApJS/209/28/table2 &   [2]       \\
B331 SW  & $ 240.03\times10^3 $ & $ 1.25  $  &  $328 \pm 14\times 10^{-3}   $ & UKIRT/WFCAM:J      & J/ApJS/209/28/table2 &   [2]       \\

\end{tabular}
\tablebib{[1]\cite{gaiacollaboration2016,gaiacollaboration2023}; [2]\cite{king2013}; [3]\cite{chambers2016}; [4] \cite{skrutskie2006}; [5] \cite{broos2013}.}
\end{minipage}
\label{tab:photometry_new_dets}
\normalsize
\end{table}

\end{appendix}
\end{document}